\documentclass[aps,prd,twocolumn,notitlepage,10pt,
nofootinbib,nobibnotes,superscriptaddress,preprintnumbers]{revtex4-1}

\usepackage{graphicx,xcolor}
\usepackage{physics}
\usepackage{amsmath,amssymb}
\usepackage{enumitem}
\usepackage{dcolumn}
\usepackage[hidelinks]{hyperref}
\usepackage[normalem]{ulem}

\usepackage{lipsum}

\newcommand{\avg}[1]{\left\langle#1\right\rangle}

\newcommand{\GeV}{\mathrm{GeV}}
\newcommand{\fm}{\mathrm{fm}}
\newcommand{\epsth}{\ensuremath{\varepsilon_\mathrm{th}}}
\newcommand{\epssw}{\ensuremath{\varepsilon_\mathrm{sw}}}
\newcommand{\sqrtsnn}{\ensuremath{\sqrt{s_{\rm NN}}}}
\newcommand{\GeVfm}{\GeV/\fm}
\newcommand{\bfr}{\mathbf{r}}
\newcommand{\bfu}{\mathbf{u}}

\begin{document}
\title{Pushing the hybrid approach to low beam energies with dynamic initial conditions from hadronic transport}
\author{Renan Góes-Hirayama}     
\affiliation{Helmholtz Forschungsakademie Hessen für FAIR (HFHF), GSI Helmholtzzentrum für Schwerionenforschung, Campus Frankfurt, Max-von-Laue-Str. 12, 60438 Frankfurt am Main, Germany
}%
\affiliation{Frankfurt Institute for Advanced Studies, Ruth-Moufang-Strasse 1, 60438 Frankfurt am Main, Germany}
\affiliation{Institut f\"{u}r Theoretische Physik, Goethe Universit\"{a}t, Max-von-Laue-Strasse 1, 60438 Frankfurt am Main, Germany}

\author{Joscha Egger}
\affiliation{Institut f\"{u}r Theoretische Physik, Goethe Universit\"{a}t, Max-von-Laue-Strasse 1, 60438 Frankfurt am Main, Germany}

\author{Zuzana Paulínyová}
\affiliation{Frankfurt Institute for Advanced Studies, Ruth-Moufang-Strasse 1, 60438 Frankfurt am Main, Germany}
\affiliation{Institut f\"{u}r Theoretische Physik, Goethe Universit\"{a}t, Max-von-Laue-Strasse 1, 60438 Frankfurt am Main, Germany}
\affiliation{Faculty of Science, P.J. Šafárik University, Šrobárova 2, 041 54 Košice, Slovakia}

\author{Iurii Karpenko}
\affiliation{Faculty of Nuclear Sciences and Physical Engineering, Czech Technical University in Prague, Břehová 7, 11519 Prague 1, Czech Republic}

\author{Hannah Elfner}%
\affiliation{GSI Helmholtzzentrum f\"{u}r Schwerionenforschung, Planckstr. 1, 64291 Darmstadt, Germany}
\affiliation{Helmholtz Forschungsakademie Hessen für FAIR (HFHF), GSI Helmholtzzentrum für Schwerionenforschung, Campus Frankfurt, Max-von-Laue-Str. 12, 60438 Frankfurt am Main, Germany
}%
\affiliation{Frankfurt Institute for Advanced Studies, Ruth-Moufang-Strasse 1, 60438 Frankfurt am Main, Germany}
\affiliation{Institut f\"{u}r Theoretische Physik, Goethe Universit\"{a}t, Max-von-Laue-Strasse 1, 60438 Frankfurt am Main, Germany}

\date{\today}

\begin{abstract}
While hybrid approaches of relativistic hydrodynamics+transport have been well established for the dynamical description of heavy-ion collisions at high beam energies, moving to lower beam energies is challenging. In this work, we propose dynamic initial conditions for the viscous hydrodynamic evolution in heavy-ion collisions at low to intermediate beam energies. They are comprised of core hadrons based on the local energy density during the pre-equilibrium hadronic evolution. The SMASH-vHLLE hybrid approach is then applied to lower beam energies, achieving good agreement with measured bulk observables between $\sqrtsnn=3$ and $9.1\ \GeV$, thus providing guidance for measurements in STAR-BES and CBM at FAIR.
\end{abstract}

\maketitle

\section{Introduction}

The main purpose of heavy-ion collision experiments is to explore the phase diagram of quantum chromodynamics (QCD). At zero baryochemical potential, lattice QCD results \cite{Bazavov:2017dus,HotQCD:2018pds} indicate that this phase diagram exhibits a smooth crossover at a temperature $T\approx156\ \mathrm{MeV}$, which connects the hadronic and deconfined degrees of freedom, and can be probed by heavy-ion collisions at very high beam energies. Recent years have seen a large experimental effort to study less energetic beams, such as the energy scans done by the STAR \cite{STAR:2010vob,STAR:2017sal} and NA61/SHINE collaborations \cite{NA61SHINE:2023epu,NA61SHINE:2025whi}, as well as in the future CBM experiment at FAIR \cite{CBM:2016kpk}. Such collisions probe a region of the QCD phase diagram inaccessible by lattice methods, lower in temperature and higher in baryon number, and may help to locate the potential first-order phase transition and critical endpoint (CEP). Detailed dynamical models are essential to extract QCD properties from these collisions.
 
A very successful framework for describing high energy heavy-ion collisions is viscous hydrodynamics in combination with transport approaches, a so called hybrid approach. The state-of-the-art picture consists of a pre-equilibrium stage after the collision of highly Lorentz-contracted nuclei, followed by a hydrodynamic evolution typically initialized at a fixed hyperbolic time $\tau=\sqrt{t^2-z^2}$. Eventually, the fireball expansion dilutes the medium enough to be described by individual hadrons, which scatter until freezeout within hadronic transport. When the beam energy is not high enough ($\sqrtsnn\leq30\ \GeV$), the picture of pancake-like ions breaks down. The finite longitudinal extent of the initial state means that (1) boost invariance no longer holds, so the evolution framework must be three-dimensional, and (2) different parts of the nuclei will interact at different times. This implies that, when a hydrodynamic description is employed, nucleons at the ``front'' form the fluid earlier than the ones at the ``back''. On top of that, there is no guarantee that the entirety of the medium approaches equilibrium, such that hydrodynamics is only applicable for the regions that do so. A proper dynamical evolution model thus needs to account for this separation. Furthermore, hybrid models have direct control over the equation of state, which provides a useful baseline on which effects of a phase transition can be studied.

Dynamical creation of a fluid at low or intermediate energies has been previously modeled in \cite{Shen:2017bsr} by decelerating the endpoints of the strings (i.e. color flux tubes) formed between colliding nucleons before fragmentation, with some additional rapidity loss fluctuation. Similarly, nucleons in \cite{De:2022yxq} lose rapidity based on binary collisions, and become hydrodynamical sources of energy and baryon charge at a given time after the last collision. The initial state model in \cite{Du:2018mpf} also uses a fixed time condition, but based on the formation time of string fragmentation products from UrQMD collisions. Alternatively to fixed time fluidization models, a hybrid model was assembled with the JAM hadronic transport and ideal hydrodynamics in \cite{Akamatsu:2018olk}, using the hadronic sources based on the local energy density of the medium. 

The SMASH-vHLLE hybrid \cite{Schafer:2021csj} is a modular hybrid that uses a fixed time initial condition based on hadronic distributions going down to $\sqrtsnn=4.3\ \GeV$. In this work, we extend it to further lower energies, aiming at the region of interest for the CEP -- according to recent lattice QCD, holography, and FRG studies \cite{STAR:2021fge,Fu:2023lcm,Hippert:2023bel,Borsanyi:2025dyp}. We do this with an approach similar to the JAM hybrid \cite{Akamatsu:2018olk} but with a rather distinct implementation and set of assumptions: we allow all hadrons to be sources for the hydrodynamic evolution, not only products of hadronic decays or string fragmentation, including leading hadrons; we observe a strong sensitivity on delaying fluidization based on the formation time of string products, and thus treat it as a model parameter; and we use a hydrodynamic evolution with finite viscosity, which is necessary due to the low temperatures that dominate the collisions of interest.

This paper is organized as follows. Section \ref{sec:energy_density_condition} explores the local energy density as a condition for switching to a hydrodynamic description. Section \ref{sec:hybrid} describes the conditions for particles becoming core and therefore part of the fluid or being left as corona, as well as the hybrid framework used in this work. Section \ref{sec:tuning} discusses our choice of default parameters, and section \ref{sec:bulk_results} presents a comparison of the results obtained with this new approach versus a pure hadronic cascade and the standard SMASH-vHLLE hybrid fixed time initial conditions.

\subsection{SMASH}\label{sec:SMASH}

The framework described in this paper is built on top of the hadronic transport approach SMASH (\emph{Simulating Many Accelerated Strongly-interacting Hadrons}) \cite{SMASH:2016zqf}, which provides an effective solution of the relativistic Boltzmann equation. Interactions are either binary ($1\leftrightarrow2$ and $2\leftrightarrow2$) or string fragmentation ($2\to N$), handled by PYTHIA \cite{10.21468/SciPostPhysCodeb.8,10.21468/SciPostPhysCodeb.8-r8.3}. Each hadron-hadron collision is determined by a geometric interpretation of the total cross section, which matches elementary cross sections when they are known, or is rescaled from the nucleon-nucleon value with the Additive Quark Model otherwise. The outcome of each interaction is randomly selected from the available partial cross-sections.

SMASH can be initialized with a variety of setups, of which we use the following two.
 
\emph{Infinite matter}: given an initial temperature and baryochemical potential, the multiplicity and momenta of each particle species are calculated from a hadron resonance gas and sampled from Boltzmann distributions in a periodic box. To let the system relax to global equilibrium hadrons propagate and interact, before the results are evaluated. If the volume and time are large enough, this setup effectively simulates a thermalized infinite hadron gas, typically with a slightly smaller temperature than the input \cite{Rose:2020lfc}.

\emph{Collider}: given two colliding nuclei, beam energy, and impact parameter, SMASH places the initial nucleons based on the Woods-Saxon distribution such that the first nucleon-nucleon collision happens at approximately $t=0$. In this work, Fermi motion is ``frozen'', only affecting the momenta of hadrons after the interactions start, an approximation that allows us to neglect nuclear potentials. While they can be relevant at the beam energies of interest, they are numerically expensive to use, especially in addition to the hydrodynamic evolution, so we leave studying their effect for the future.

\section{Energy density as a condition for fluidization}\label{sec:energy_density_condition}

The hydrodynamic evolution in simulations of heavy-ion collisions typically starts at a constant hyperbolic time, with the specific value $\tau_0$ chosen either as the nuclear passing time \cite{Schafer:2021csj,Karpenko:2015xea} or as a result of a match to experimental data \cite{Gotz:2025wnv,Nijs:2020ors}. In low beam energy collisions, the timescale for initial nucleon-nucleon interactions is comparable to the fireball expansion, so a fixed-time initial condition for the hydrodynamic equations is unphysical. We propose the alternative condition of a minimum energy density, above which the particle description locally gives way to a fluid description. 

This fluidization condition is motivated by Ref.~\cite{Inghirami:2022afu} (see figs. 12a and 12c), where a negative correlation was found between energy density and the local deviation from equilibrium, for event-by-event simulations of heavy-ion collisions in SMASH. As we discuss in Appendix \ref{sec:mfp}, the energy density is a good probe of the mean free path in a hadron gas, meaning that an energetic region corresponds to a large interaction rate, and tends towards equilibrium faster than a dilute region \cite{Rose:2020lfc}. Fluidization is the terminology we employ to describe the conversion of particles into fluid, in analogy to the particlization that happens on the Cooper-Frye hypersurface. 

\begin{figure}[t]
\centering
\includegraphics[width=.85\linewidth]{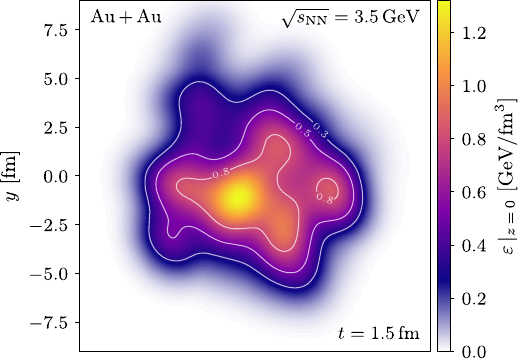}
\includegraphics[width=.85\linewidth]{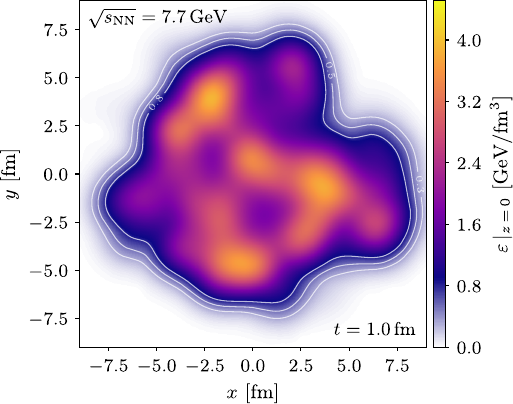}
\caption{Energy density profile of a head-on Au+Au collision in the transverse plane at $z=0$ (upper) and $t=1.5\ \fm$ at $\sqrtsnn=3.5\ \GeV$ (lower) and $t=1.0\ \fm$ at $\sqrtsnn=7.7\ \GeV$. The white lines denote $\varepsilon=0.3,0.5,$ and $0.8\ \GeVfm^3$.}\label{fig:ed-heatmap}
\end{figure}

We illustrate this condition in fig.~\ref{fig:ed-heatmap} with the energy density profile at $z=0$ for two head-on Au+Au collisions with zero impact parameter at different energies as simulated in SMASH. The upper plot is a snapshot of a $\sqrtsnn=3.5\ \GeV$ collision at $t=1.5\ \fm$, while the lower is at $7.7\ \GeV$ and $1\ \fm$, and the white curves demark lines of constant energy density at $\varepsilon=0.3,0.5,$ and $0.8\ \GeVfm^3$. 

The lines of constant energy density tend to increase in value towards the center, with fluctuations from the nuclear configuration which form hot spots, such that these lines are irregular and can even be disjointed. In the upper plot with the lower beam energy, the lowest value $\varepsilon=0.3\ \GeVfm^3$ encompasses most of the fireball, only excluding the very dilute edges, while the highest value $0.8$ contains only the dense center, where substructures of even higher $\varepsilon$ can be seen. At a higher beam energy, the curves in this $0.3-0.8\ \GeVfm^3$ range clump up at the edges, since the system is much more energetic, so the specific value for this hydrodynamization condition matters less at higher beam energies. Therefore, using a ``threshold'' energy density $\epsth$ produces a natural core-corona separation at low beam energies.




\begin{figure*}[ht]
    \centering
    \includegraphics[width=0.32\linewidth]{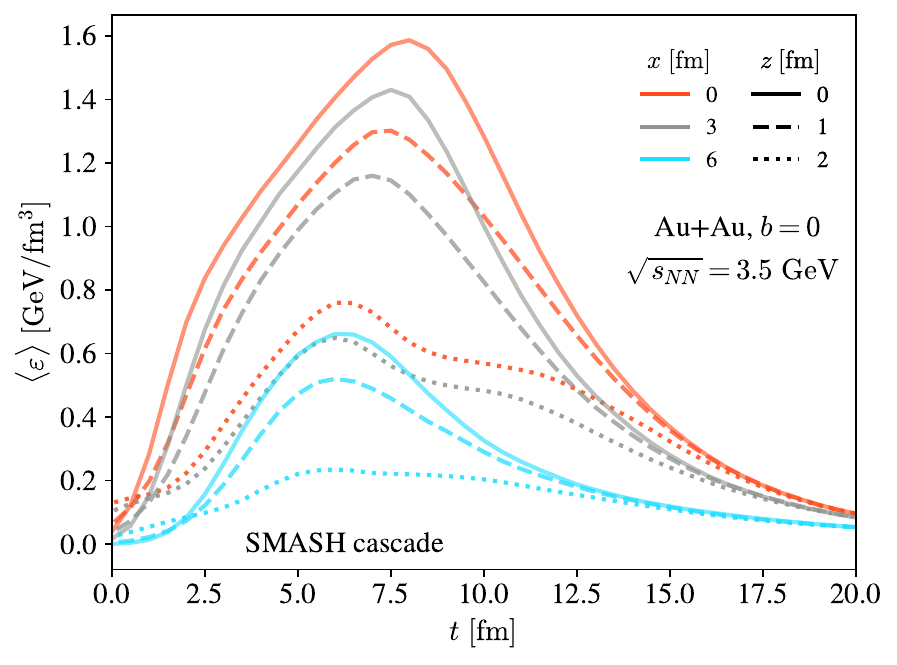} %
    \includegraphics[width=0.32\linewidth]{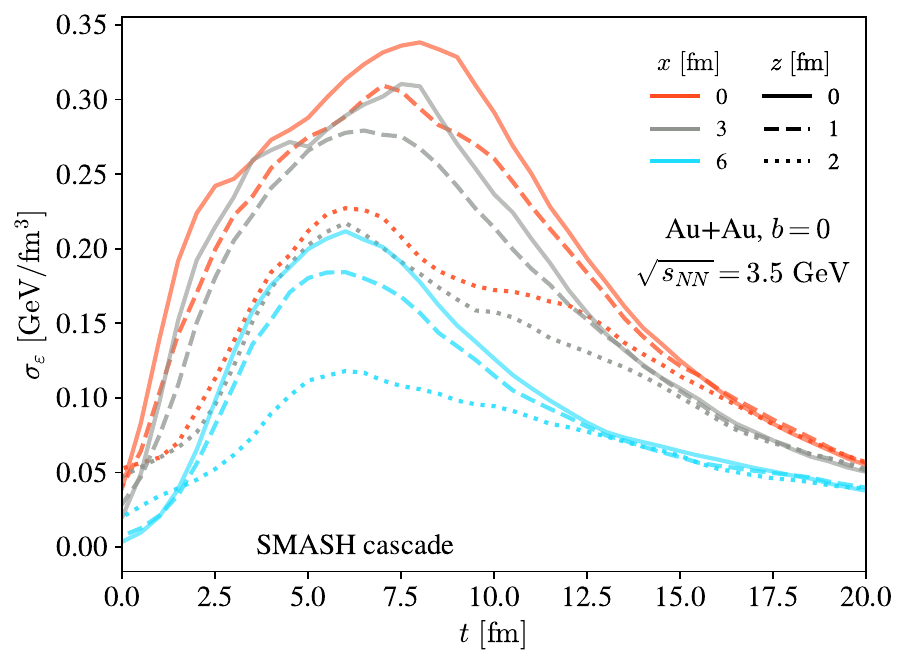}
    \includegraphics[width=0.32\linewidth]{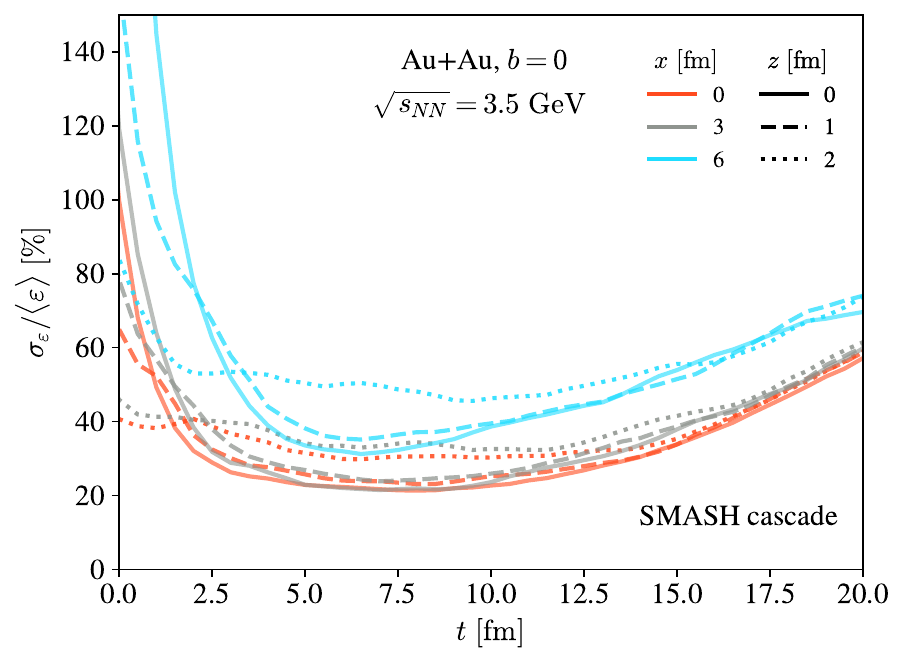}
    \caption{Evolution of (left) the mean energy density $\avg{\varepsilon}$, (middle) its event-by-event fluctuation $\sigma_\varepsilon$, and (right) the ratio $\sigma_\varepsilon/\avg{\varepsilon}$ in different spatial points (with $y=0$) for a head-on Au+Au collision at $\sqrtsnn=3.5\ \GeV$.}\label{fig:evol_energydensity}
\end{figure*}

In fig. \ref{fig:evol_energydensity}, we show the evolution of the average energy density and its event-by-event fluctuations for the $\sqrtsnn=3.5\ \GeV$ collision, where each curve tracks a point in space. The left plot shows that points closer to the center have higher average values, as seen in fig. \ref{fig:ed-heatmap}, reaching the maximum at $x=y=z=0$ around $t=\tau_0=8.1\ \fm$, with $\tau_0$ given by eq. \eqref{eq:constant_tau}. In the other regions, maxima are smaller and reached earlier, since less matter overall passes through them. Each curve crosses a given $\avg{\varepsilon}$ in different times or does not cross at all, even for the same $z$. This highlights that, if a threshold $\epsth$ determines whether this region can be described hydrodynamically, the usual fixed-time initial conditions are not appropriate. 

The evolution of event-by-event fluctuations of the energy density in the middle and right plots further show that parts of the system may exceed a given $\epsth$ only in some events, in particular close to the edge of the fireball. For example, the average energy density at $(x,y,z)=(6,0,0)\ \fm$ point never reaches $\epsth=0.8\ \GeVfm^3$, but since $\varepsilon+\sigma_\varepsilon>0.8\ \GeVfm^3$ at the maximum, a non-negligible fraction\footnote{ $P(\varepsilon>\avg{\varepsilon}+\sigma_\varepsilon)=15.8\%$, assuming the distribution is normal.} of the events could be described as a fluid. This is a consequence of the aforementioned hotspots, suggesting that an average initial condition does not capture important physics, as these hotspots introduce large pressure gradients in the hydrodynamic fields. Because the relative fluctuations of energy density are not negligible, we produce only event-by-event calculations.

We further note the findings in Ref. \cite{Inghirami:2022afu}, where better suited conditions for the applicability of hydrodynamics, based on the off-diagonality and anisotropy of the energy-momentum tensor, were imposed for collisions between $\sqrtsnn=2.4-7.7\ \GeV$. It was found that these conditions are only fulfilled in a significant part of the evolution for averaged events, and rarely in an event-by-event basis. However, the correlation between high energy density volume elements and closeness to an equilibrium energy-momentum tensor vanishes in averaged initial states. 





\section{Hybrid framework}\label{sec:hybrid}

This work was done using the SMASH-vHLLE hybrid approach. SMASH is employed to sample the nucleons according to Woods-Saxon distributions and calculate the microscopic collision dynamics to provide a set of hadrons to serve as initial condition. The viscous hydrodynamic code vHLLE \cite{Karpenko:2013wva,Huovinen:2012is} is used to evolve the core until a predefined switching energy density, when a particlization hypersurface is constructed. The SMASH-hadron-sampler \cite{Karpenko:2015xea} samples hadrons according to the Cooper-Frye procedure, feeding them back into SMASH for the late stage hadronic rescattering. The communication between the codes is done by the Hybrid-handler \cite{sciarra_2025_15880337}, a convenient tool for hybrid approaches. The specific branches used are publicly available in \cite{branchHybrid}.
This work was done using the SMASH-vHLLE hybrid approach. SMASH is employed to sample the nucleons according to Woods-Saxon distributions and calculate the microscopic collision dynamics to provide a set of hadrons to serve as initial condition. The viscous hydrodynamic code vHLLE \cite{Karpenko:2013wva,Huovinen:2012is} is used to evolve the core until a predefined switching energy density, when a particlization hypersurface is constructed. The SMASH-hadron-sampler \cite{Karpenko:2015xea} samples hadrons according to the Cooper-Frye procedure, feeding them back into SMASH for the late stage hadronic rescattering. The communication between the codes is done by the Hybrid-handler \cite{sciarra_2025_15880337}, a convenient tool for hybrid approaches. The specific branches used are publicly available in \cite{branchHybrid}.

In the default hybrid configuration, the initial condition is based on a fixed hyperbolic time, set to the \emph{nuclear passing time}
\begin{equation}\label{eq:constant_tau}
    \tau_0 = \frac{2m_\mathrm{N}}{\sqrt{s_\mathrm{NN}-4m_\mathrm{N}^2}}(R_\mathrm{P}+R_\mathrm{T}),
\end{equation}
where $m_\mathrm{N}$ is the nucleon mass, $\sqrtsnn$ the center-of-mass energy between a nucleon pair, and $R_\mathrm{P(T)}$ is the radius of the projectile (target) nucleus. When a hadron crosses $\tau_0$, it is fed to vHLLE and removed from the evolution, so we call this initial condition ``iso-$\tau$''.

Ideally, a truly dynamical evolution would require running both off-    equilibrium hadronic transport and viscous hydrodynamics concurrently, which is technically challenging. We restrict ourselves to a simpler scenario, where core hadrons in the hadronic cascade act as proxies for the forming fluid. The selection of core hadrons and the dynamical initial conditions, which are the major aim of this work, are described in more detail below.  

\subsection{Core-corona separation}

\begin{figure*}[ht]
\centering
\includegraphics[width=\linewidth]{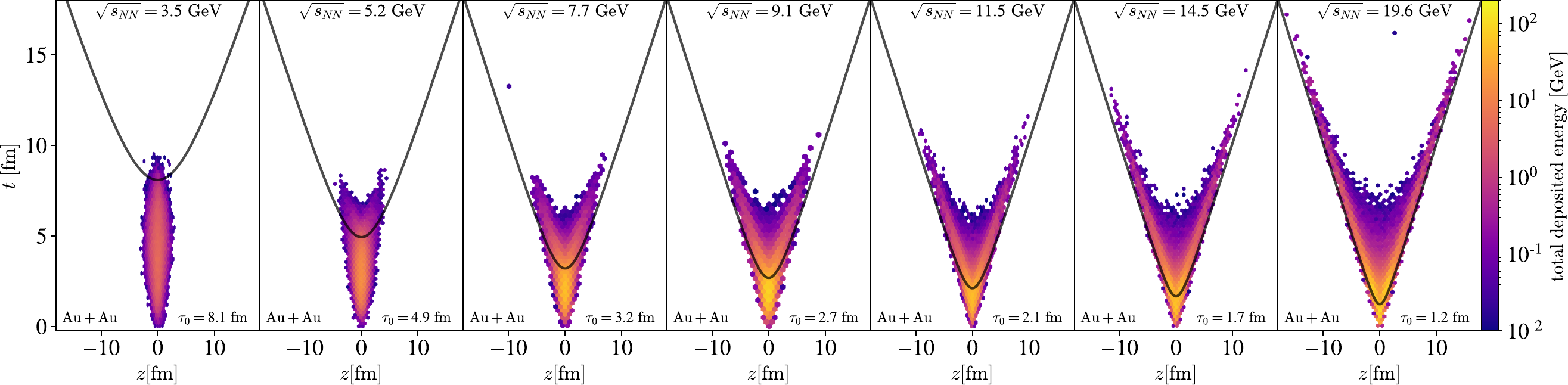}\caption{Spacetime distribution of energy deposition within our core-corona separation for head-on Au+Au collisions at different beam energies, using $\epsth=0.5\ \GeVfm^3$ and $t_f=0.25\ \fm$. The black hyperbolas show the constant hypersurface of constant $\tau$ defined by the nuclear passing time $\tau_0$ \eqref{eq:constant_tau}.}\label{fig:fluidize_spacetime}
\end{figure*} 

Based on the distribution function that is propagated in SMASH by testparticles a coarse lattice is constructed, where local densities are computed from the hadrons. The energy-momentum tensor is evaluated in a lattice cell in the lab frame as
\begin{equation}\label{eq:hadron_Tmunu}
    T^{\mu\nu}(\mathbf{r}) = \sum_i \frac{p_i^\mu p_i^\nu}{p_i^0}K\left(\mathbf{r}-\mathbf{r}_i,\frac{\mathbf{p}_i}{m_i}\right),
\end{equation}
where $\mathbf{r}_i$, $p^\mu_i$, and $m_i$ are respectively the position, 4-momentum, and pole mass of particle $i$. The energy-momentum tensor is smoothed by a covariant smearing kernel normalized to unity,
\begin{equation}\label{eq:smearing_kernel}
    K(\bfr,\bfu)=\frac{\gamma}{(2\pi\sigma^2)^{3/2}}\exp\left(-\frac{|\bfr|^2+(\bfr\cdot\bfu)^2}{2\sigma^2}\right),
\end{equation}
with the Lorentz factor $\gamma$. 

The basic condition for particle $j$ to fluidize (become part of the core) is that the local energy density exceeds a predefined threshold: 
\begin{equation}\label{eq:dynflu_threshold}
    T^{00}_\mathrm{LRF}(\mathbf{x}_j) - \frac{m_j}{(2\pi\sigma^2)^{3/2}}  \geq \epsth,
\end{equation}
in which $T^{\mu\nu}_\mathrm{LRF}$ is \eqref{eq:hadron_Tmunu} boosted to the local (Landau) rest frame of the cell containing $j$, and the energy density threshold $\epsth$ is a model parameter. The mass term accounts for the own contribution -- a very massive particle should not become fluid by itself.

Condition \eqref{eq:dynflu_threshold} is necessary but not sufficient for fluidization, another constraint is imposed to prevent unphysical behavior. In SMASH, string fragmentation is implemented as an interaction at a fixed point in spacetime. Since string production constitutes the major part of the cross section above $\sqrtsnn\geq4\ \GeV$, relying only on high energy densities fluidizes most of the system instantly in extreme peaks that are inconsistent with hydrodynamics. We introduce the \emph{string fluidization time} $t_f$ as one more model parameter, corresponding to the minimum time that needs to pass in the rest frame of a string fragmentation product before \eqref{eq:dynflu_threshold} is evaluated. $t_f$ is akin to the formation time of hadrons, but smaller, as the deposition of energy into the fluid could come from partons settling into the QGP. It is important to note that, for low beam energy collisions in SMASH, the string fragmentation products can interact before the formation time, but do so with reduced cross sections \cite{Mohs:2019iee}. To deter extreme and instantaneous hotspots, we also forbid the fluidization of initial nucleons that did not scatter, or scattered elastically only once.

A particle fulfilling the conditions above is tagged as \emph{core}. When this happens, the relevant quantities (e.g. position, momentum, quantum numbers) of the particle are written to output, which serves as the input for the subsequent hydrodynamic calculation. The decay products of core resonances are created as part of the core and not written to output, as this would double count the deposition into the fluid. Core particles are kept in the evolution and retain their interactions with each other until the end of this stage, which is necessary to prevent against a fully transparent core. However, the only cross section between core and corona is elastic, which avoids the ambiguity in deciding which part of the outgoing state was already in the core. These elastic interactions can transfer energy between core and corona, but we assessed that the net loss of energy is at most 2\% at $\sqrtsnn=3\ \GeV$, and 1\% at $\sqrtsnn=19.6\ \GeV$. This initial stage is evolved until $t=20\ \fm$, enough for the core to be completely selected in the collisions of interest, and the corona particles are backpropagated to the 4-position where they last interacted.

This core-corona separation leads to fig. \ref{fig:fluidize_spacetime}, which shows how energy from fluidized hadrons is deposited into the core across spacetime (integrated in the transverse direction) for the intermediate range of beam energies. In the lower end, most of the system reaches the fluidization threshold substantially earlier than the traditional iso-$\tau$ initialization that is depicted as the black hyperbolas. Because the colliding nucleons are largely stopped, the deposition exhibits no structures along the beam direction $z$. When the beam energy increases, however, the nucleons pass more through each other, fluidizing the produced medium in more forward regions. At high enough energies, most of the deposited energy coincides with the $\tau_0$ hyperbola, with a tail that increases with $t_f$. In contrast to the iso-$\tau$ initialization, we name the current procedure \emph{Dynamic Fluidization} (DynFlu).

As a technical note, while SMASH does not require time steps to run, the lattice where we evaluate \eqref{eq:hadron_Tmunu} and \eqref{eq:dynflu_threshold} is updated at fixed intervals of $\Delta t=0.05\ \fm$, with cubic cells with side length $\Delta x_i=0.5\ \fm$. 

\subsection{Hydrodynamic evolution}

The hydrodynamic evolution is performed in a Cartesian version of the viscous 3+1-dimensional hydrodynamic code vHLLE \cite{Karpenko:2013wva,Karpenko:2015xea}. The original version of the code is intended for modeling nuclear collisions at very high energies using hyperbolic coordinates in $t-z$ plane, $z$ being the collision axis. However, for our present application in dynamic fluidization, the Cartesian coordinate frame is a better choice: first, at low collision energies, there is no boost invariance along the longitudinal direction $z$; second, SMASH runs in Cartesian coordinates and for seamless communication between the two approaches, the same coordinate system is required. 

In addition, the option to melt hadrons into the hydrodynamic evolution throughout its duration has been added to vHLLE to allow for dynamic fluidization. The list of core hadrons from SMASH is used as a queue of particles waiting to to be melted into the fluid. The corresponding source terms for the energy-momentum and charge conservation equations from the fluidizing hadrons are as follows:
\begin{equation}\label{eq:source_currents}
\begin{aligned}
    j^\mu(\mathbf{r})&=\sum_i p^\mu K\left(\mathbf{r}-\mathbf{r}_i,\frac{\mathbf{p}_i}{m_i}\right), \\
    j_C(\mathbf{r})&=\sum_i C_i K\left(\mathbf{r}-\mathbf{r}_i,\frac{\mathbf{p}_i}{m_i}\right),
\end{aligned}
\end{equation}
where $p^\mu$ is the energy, momentum and $C_i$ are the baryon and electric charges of hadron $i$ and $K$ is given by \eqref{eq:smearing_kernel}, for consistency with SMASH. The initial condition for hydrodynamics is constructed from the particles entering at the earliest time. The hydrodynamic evolution starts with this initial state and the next particles are melted when the calculation time passes their fluidization time, taking the form of source terms in the hydrodynamic equations of motion for both $T^{\mu\nu}$ and $j^\mu$:
\begin{equation}
\begin{aligned}
 \partial_\nu T^{\mu\nu} &= j^\mu(\mathbf{r}), \\
 \partial_\nu N^\nu_C &= j_C(\mathbf{r}),
\end{aligned}
\end{equation}

The hybrid approach with iso-$\tau$ initial state was tuned  with a Bayesian analysis performed in \cite{Gotz:2025wnv}. We use the maximum a posteriori (MAP) parameters from this analysis for the fluid evolution. The shear viscosity uses two straight lines as a function of temperature, notably equal to 0 for high temperatures if the MAP values are used. The bulk viscosity is parametrized as a function of the energy density with an asymmetric Gaussian. The specific functional forms and parameters are detailed in appendix~\ref{sec:visc_parameters}. Charge diffusion currents are not included in the fluid dynamic equations as we leave this extension for a future update.

We use a chiral mean-field equation of state \cite{Motornenko:2019arp,Most:2022wgo}, which has no phase transition, matched to a hadron resonance gas at lower energy densities. In the future, changing the equation of state will present an opportunity to systematically study the effects of a phase transition onto final-state observables in a state of the art evolution framework. 

\subsubsection*{Particlization and late-stage}

A particlization hypersurface is constructed via the Cornelius routine \cite{Huovinen:2012is} during the hydrodynamical evolution. For every hypercube containing 16 neighboring fluid cells in space and time, an element of particlization surface is found once some cells have their rest frame energy density above $\epssw$ and some below $\epssw$. The surface elements found are stored in an output file. While naively one might expect $\epssw$ to be equal to $\epsth$, this is not necessarily the case, as discussed in Appendix \ref{sec:fluidization_parameters}. Instead, we use $\epssw=0.334\ \GeVfm^3$ as determined in \cite{Gotz:2025wnv}.

The evolution stops once all fluid cells have crossed $\epssw$. If this happens before all sources enter the fluid, the remaining particles are regarded as part of the corona. This situation occurs for a small number of hadrons in a few events, particularly at the lowest energies, where the hydrodynamic stage is rather short-lived. In the beginning of the expansion, however, this poses a particular problem in the DynFlu scenario, where the small bubbles of the fluid formed by a few initial particles are quick to expand and cool down, possibly stopping the simulation prematurely. We circumvent this by postponing the hypersurface construction until enough sources have entered the fluid. In our current setup, 15 particles suffice to sustain a stable hydrodynamic evolution. 

The hypersurface information is used by the SMASH-hadron-sampler \cite{Karpenko:2015xea} for particlization according to the Cooper-Frye prescription, including shear and bulk viscous corrections \cite{Denicol:2014vaa,Paquet:2015lta} to sample hadrons from a thermal distribution. The Cooper-Frye formula assumes a non-interacting gas of stable hadrons and hadronic resonances, whereas the EoS in the fluid-dynamic evolution includes contributions from mean field potentials at non-zero baryon density, and a list of hadron states which is smaller from the list in SMASH. To ensure consistency between the two, 
temperature, chemical potentials and flow velocities are recomputed using the SMASH hadron gas equation of state. The condition of continuity of energy and momentum flow through the surface is used in the recomputation procedure.

The sampled particle list is then combined with the corona particle list of hadrons that either never fluidized or were flagged as core particles in SMASH but did not enter the fluid. The joint particle list is then processed in SMASH to model non-equilibrium final state hadronic interactions in the so-called afterburner.

\subsubsection*{Energy deposition and trajectories in the phase diagram}

\begin{figure}[t]
\centering
\includegraphics[width=0.92\linewidth]{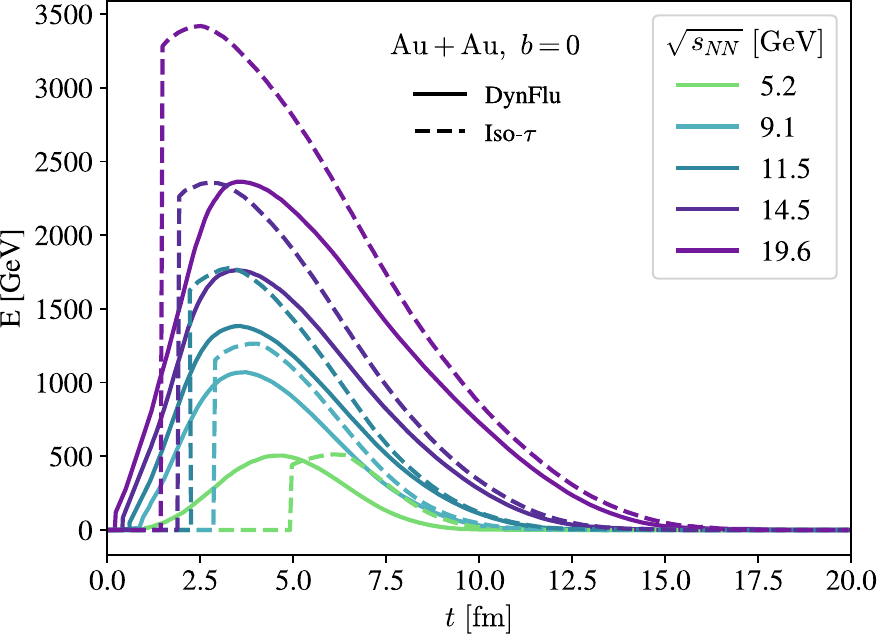}\caption{Evolution of the total energy present in the fluid during the hydrodynamic stage, initialized with the dynamic (full lines) vs. iso-$\tau$ (dashed lines) initial conditions.}\label{fig:evolution_hydro_energy}
\end{figure}

To illustrate the difference between feeding the fluid with dynamical and iso-$\tau$ sources, fig. \ref{fig:evolution_hydro_energy} shows the evolution of the total energy contained in the hydrodynamic evolution. 

Consistent with fig. \ref{fig:fluidize_spacetime}, the deposition of energy starts early on, and smoothly peaks between $t=3-5\ \fm$, depending on the collision energy. This is in contrast to the iso-$\tau$ deposition, which starts abruptly at $t=\tau_0$ and peaks a bit after that, when particles at $z>0$ cross the hypersurface. Because the dynamic fluidization is stretched over a longer interval, some regions of the fluid fall below the energy density $\epssw$ at which particlization takes place, before all sources are introduced. Consequently, the peak total energy is lower than in the iso-$\tau$ picture, where most of the energy is deposited as the hydrodynamic evolution starts. Despite this, after the peak the curves for each collision energy approach each other, because the sources in DynFlu continue to enter the fluid, slowing down the depletion. Even if the final conditions are similar, the drastically different initial dynamics affect some observables, such as electromagnetic probes, which are emitted throughout the evolution.

\begin{figure}[t]
\centering
\includegraphics[width=0.92\linewidth]{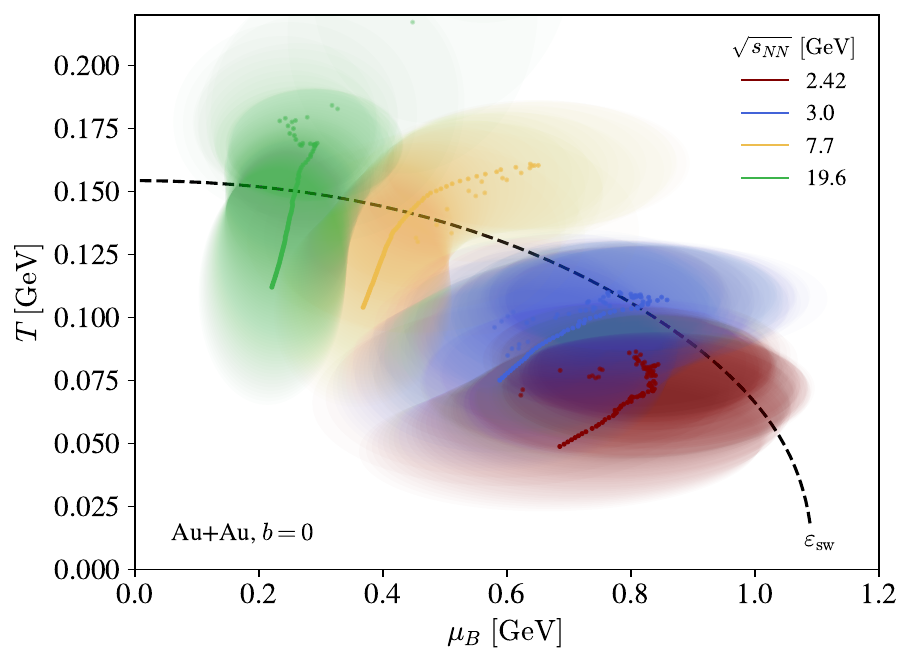}\caption{Trajectories of the fluid in the phase diagram. Each dot is an average value $\avg{\mu_B}_\varepsilon$ and $\avg{T}_\varepsilon$, and the ellipsis around each point represents their 1$\sigma$ spread (see text). The dashed black curve shows the particlization at $\epssw=0.334\ \GeVfm^3$.}\label{fig:Hydro_TmuB_phasediagram}
\end{figure}

It is also instructive to look at the path taken by the hydrodynamic evolution in the phase diagram. Figure \ref{fig:Hydro_TmuB_phasediagram} shows the trajectories for a single head-on Au+Au collision across the low-to-intermediate beam energies. The values are computed as spatial averages weighted by energy density with
\begin{equation}\label{eq:averaging}
\avg{A}_\varepsilon = \frac{\int\dd^3x~\varepsilon(x)A(x)}{\int\dd^3x\ \varepsilon(x)},
\end{equation}
where the integral is realized as a sum over all cells. Each point corresponds to the averages $\avg{\mu_B}_\varepsilon$ and $\avg{T}_\varepsilon$ at some time $t$. The ellipse surrounding it represents the second central moment of the spatial distribution, with the standard deviation as the axes sizes and the covariance $\avg{\mu_BT}_\varepsilon-\avg{\mu_B}_\varepsilon\avg{T}_\varepsilon$ determining the angle. The interval between points is $\Delta t=0.1\ \fm$ and the opacity increases when the ellipses overlap, so the opaque regions are where the system spends a longer time.

Naturally, increasing the beam energy moves the phase diagram trajectory from low $T$ and high $\mu_B$ to a region with high $T$ and low $\mu_B$. Each trajectory covers about $\Delta T=0.1\ \GeV$ and $\Delta \mu_B=0.15\ \GeV$ at particlization\footnote{In our EoS tables $\mu_B$ does not exceed $1\ \GeV$. The right ends of the red and blue ellipses are artifacts from plotting only the 2\textsuperscript{nd} moments.}.

Fluid cells with rest-frame energy density below $\varepsilon_{\rm sw}$ have already passed through the particlization surface, however their content is not drained and continues to be a part of the fluid corona. Since vHLLE only stops when all cells have $\varepsilon<\epssw$, the average trajectories extend far past it. This is more pronounced for lower energies, notably $\sqrtsnn=2.42\ \GeV$, where all of the averages lie below the hypersurface and the fluid description only applies to the hotspots. We do not expect the bulk observables to be affected by the system only containing small blobs of fluid therefore, we focus on beam energies from $\sqrtsnn = 3\ \GeV$ and above.

\section{Parameter sensitivity}\label{sec:tuning}
\begin{figure*}[ht]
    \centering
    \includegraphics[width=0.32\linewidth]{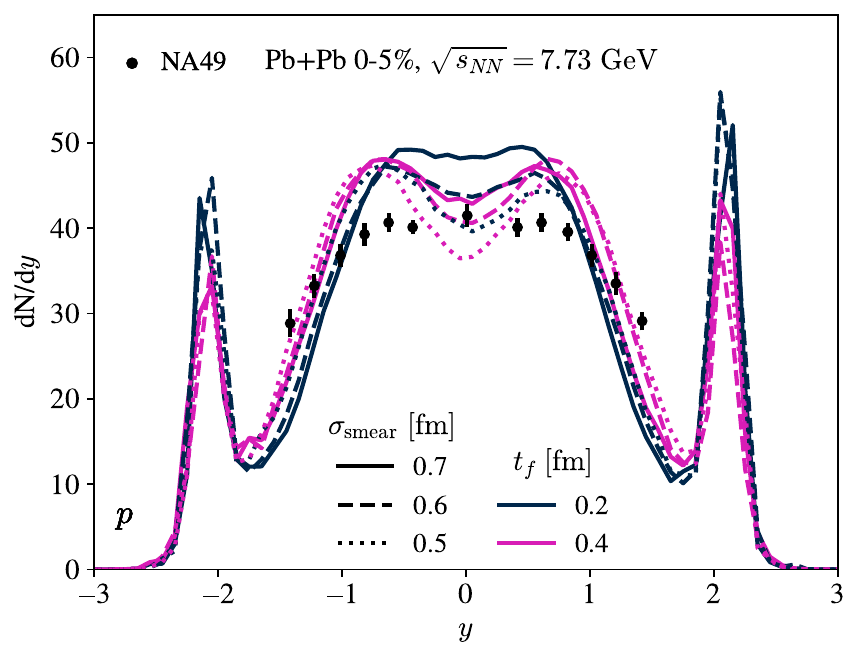}%
    \includegraphics[width=0.32\linewidth]{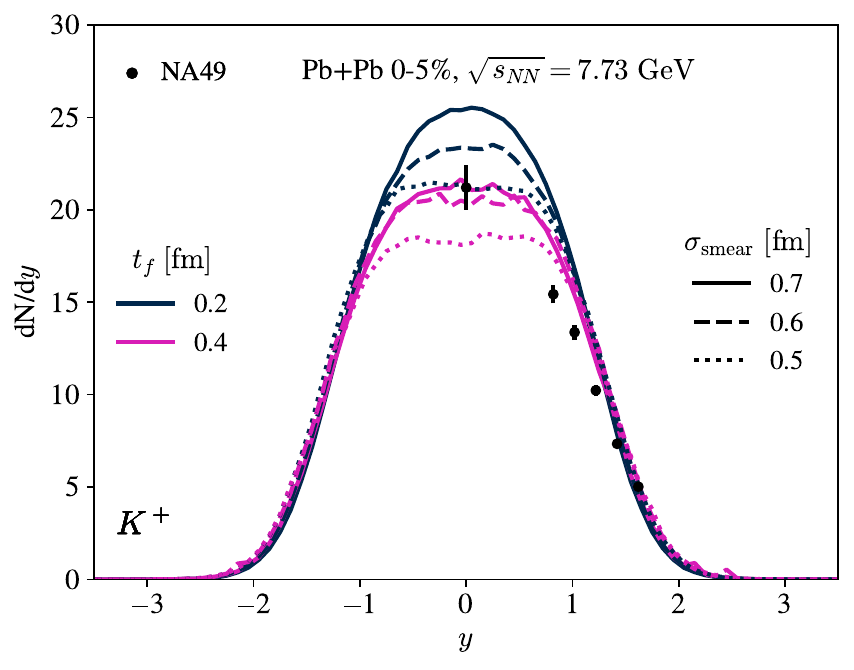}%
    \includegraphics[width=0.33\linewidth]{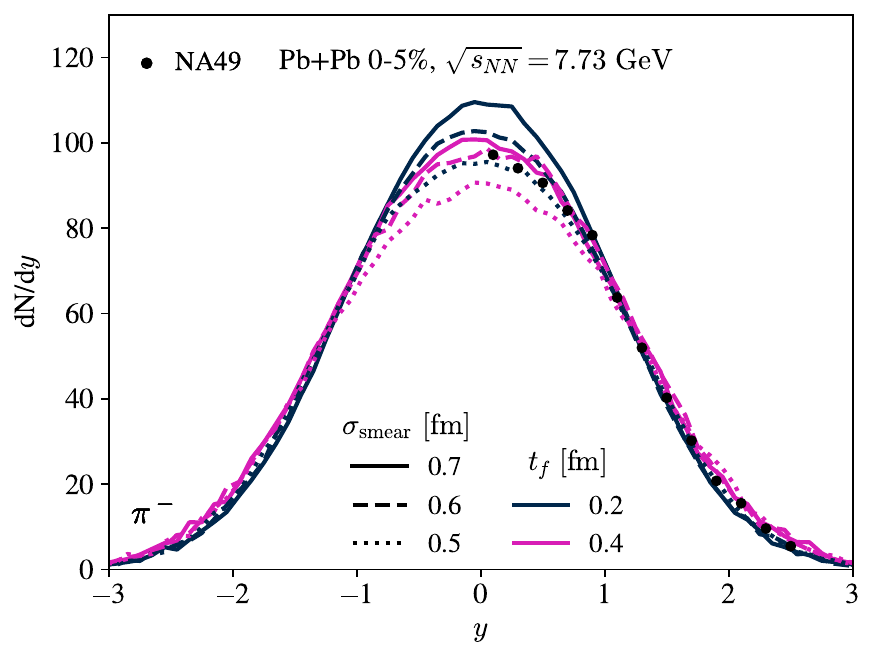}
    
    \includegraphics[width=0.32\linewidth]{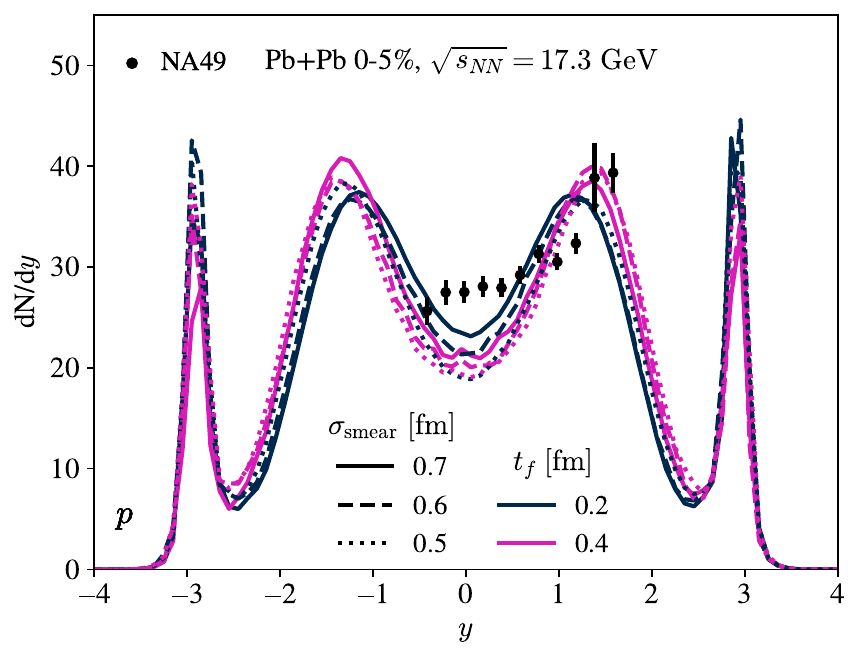}%
    \includegraphics[width=0.32\linewidth]{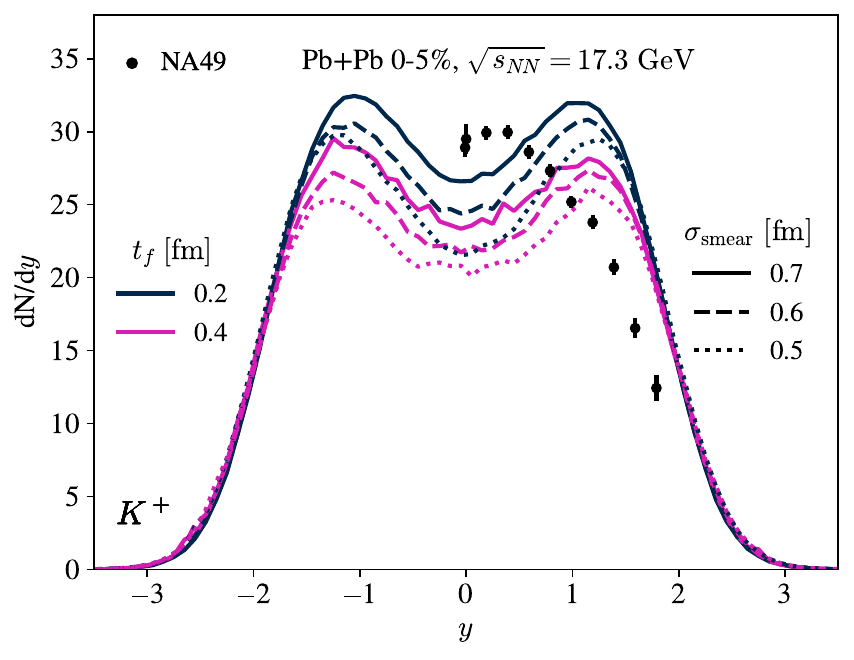}%
    \includegraphics[width=0.33\linewidth]{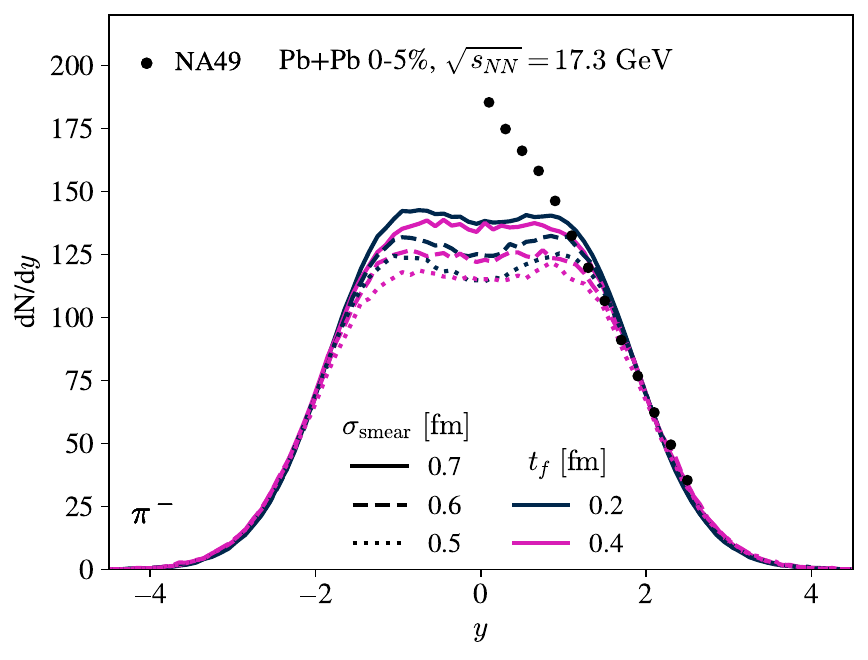}
    \caption{Rapidity distributions of (left) protons, (middle) kaons, and (right) pions for Pb+Pb collisions at (upper) 7.7 and (lower) 17.3 GeV, for varying string fluidization time $t_f$ and smearing parameter $\sigma$. Data points are from NA49 \cite{Blume:2004ci}.
    }\label{fig:tuning_smearing}
\end{figure*}

Hybrid approaches rely on different appromixations for the different parts of the evolution and therefore, some parameters to describe the interfaces are indispensible. In this work, as we aim to introduce and explore the qualitative aspects of a dynamic fluidization approach, we do not perform a rigorous tuning of \emph{all} parameters, and focus on the specific parameters threshold energy density $\epsth$, string fluidization time $t_f$, and gaussian smearing width $\sigma$. In appendix \ref{sec:fluidization_parameters} we present the physical constraints $\epsth$ and $t_f$ must obey to obtain a sensible evolution. In this range, we found that the bulk observables for which data is available do not have enough sensitivity to constrain $\epsth$ further. In this section, we use $\epsth=0.5\ \GeVfm^3$. 

Recently, a Bayesian analysis was performed for the default hybrid configuration \cite{Gotz:2025wnv}, which uses the iso-$\tau$ initialization and evolves the hydrodynamics in hyperbolic coordinates. Apart from $\epsth$, $t_f$, and $\sigma$, we use the maximum-a-posteriori parameters estimated in this analysis, summarized in appendix \ref{sec:visc_parameters}. 
 
We determine the Cartesian smearing parameter $\sigma$ used in the kernel \eqref{eq:smearing_kernel} by comparing rapidity distributions of identified particles from Pb+Pb collisions at $\sqrtsnn=7.73$ and $17.3\ \GeV$ to experimental data. Figure \ref{fig:tuning_smearing} shows the simultaneous dependence on $\sigma$ and $t_f$, for fixed $\epsth=0.5\ \GeVfm^3$. Centrality classes are defined by impact parameter using \cite{nuclearoverlap}, with  0-5\%  as $b\in[0,3.3]\ \fm$ in both Au+Au and Pb+Pb. Since $\sigma$ and $t_f$ are highly correlated, we first look at their individual effects before deciding on  the chosen value. 

The $t_f$ parameter effectively controls the stopping power of the initial state, as clearly seen in the rapidity distributions of protons (left of Fig. \ref{fig:tuning_smearing}), in which an increasing $t_f$ transports protons from mid to forward rapidities. In addition, a larger $t_f$ leads to fewer hydrodynamic sources and hence less entropy, decreasing the yields of kaons and pions. The smearing parameter $\sigma$ represents the width of the hadron used as a source, but without an established prescription to determine it, we leave it as a free and constant parameter, independent of the hadron species. It is also connected to the entropy jump during fluidization \cite{Petersen:2010zt}, and thus affects the particle  production, in particular kaons and pions. Moreover, a wider source contributes less to the pressure gradients in the fluid, leading to softer radial flow \cite{Holopainen:2010gz}. While we observe this softening in the simulations, the effect is not sufficient to help constrain $\sigma$, therefore this plot is omitted from the presentation. 

We observe a strong negative correlation between $t_f$ and $\sigma$. Comparison to experimental data shows that, while the $\sqrtsnn=\mathrm{7.73}\ \GeV$ beam energy is well described, there are not enough particles at midrapidity in collisions at 17.3 GeV. The pion yield in particular would require a much larger smearing, which would destroy the agreement at lower beam energies. This suggests the initial state needs more stopping power at the higher energies, or an improved smearing parametrization e.g. by making $\sigma$ depend on the hadron species used as source.

For this exploratory work, we chose the constant values $\sigma=0.7\ \fm$ and $t_f=0.4\ \fm$ that reasonably describe the data at $7.7\ \GeV$, which is in our target energy range for the dynamic fluidization model. At higher energies the default hybrid framework with iso-$\tau$ initial conditions is sufficient to provide a good description. In the following we also choose a constant value of $\epsth=0.6\ \GeVfm^3$ instead of the previously used $\epsth=0.5\ \GeVfm^3$, which as we checked, results in the same values for $\sigma$ and $t_f$.

\section{Bulk observables}\label{sec:bulk_results}

\begin{figure*}[th]
    \centering
    \includegraphics[width=0.32\linewidth]{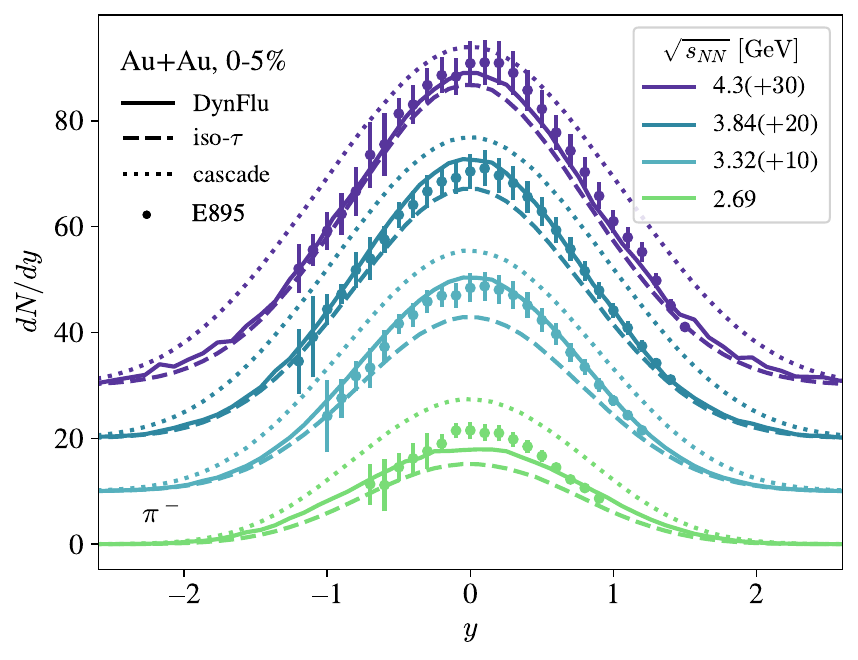}%
    \includegraphics[width=0.32\linewidth]{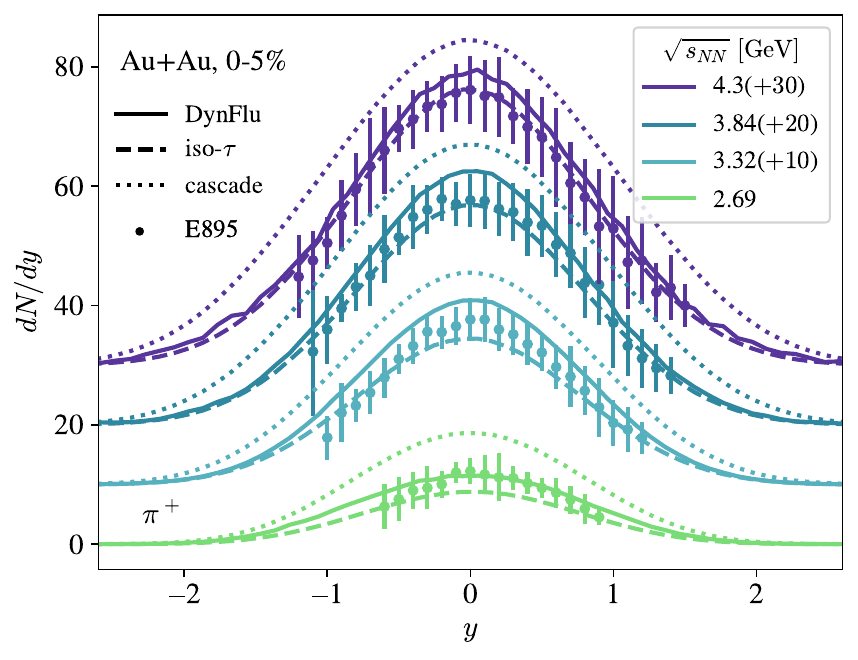}%
    \includegraphics[width=0.33\linewidth]{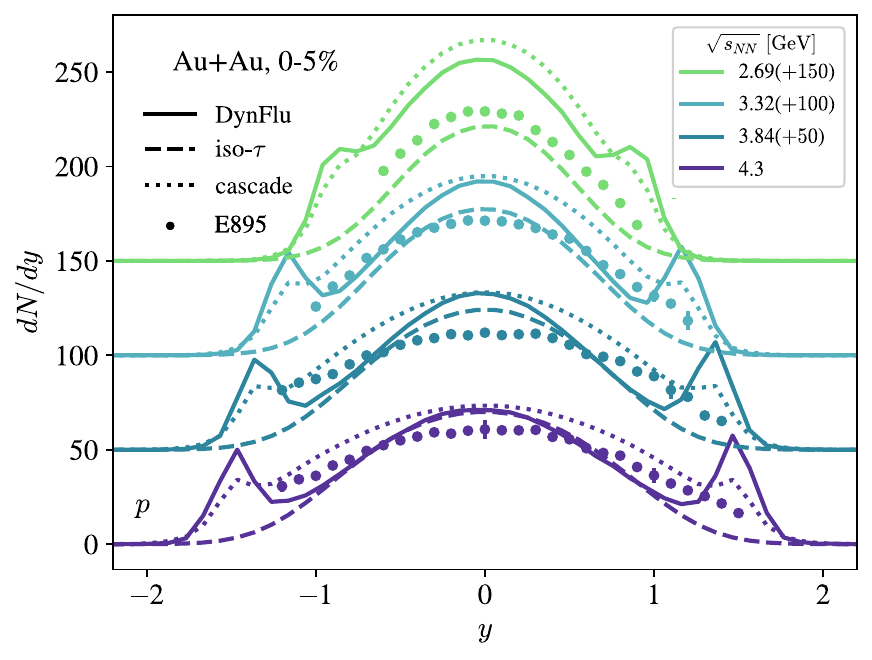}
    
    \caption{Rapidity distributions of (left) $\pi^-$, (middle) $\pi^+$, and (right) protons for central Au+Au collisions at low beam energies, comparing results from pure cascade, default iso-$\tau$ initial conditions, and dynamic fluidization. Data points are from the E895 experiment at AGS \cite{E-0895:2003oas}.
    }\label{fig:bulk_y_low_energy}
\end{figure*}

As a proof of concept, we check how the dynamic fluidization compares to the default hybrid approach based on iso-$\tau$ initial conditions and the pure hadronic cascade in SMASH, at the level of bulk observables. 

\subsubsection*{Rapidity}
We first show the rapidity distribution of protons and charged pions in AGS collisions in figure \ref{fig:bulk_y_low_energy}. The SMASH cascade overestimates the pion production at low energies. In line with was seen in \cite{Schafer:2021csj}, the usual iso-$\tau$ initialization adds \emph{all} particles into a fluid at a relatively late $\tau_0$, such that the hydrodynamic evolution is short lived and hence not much entropy can be produced, leading to a suppressed pion yield. The dynamically fluidized medium is in the middle of these descriptions, since a significant part of the system -- approximately 22\% in this setup -- stays in the corona. 
Notably, the difference between the iso-$\tau$ and DynFlu initialization decreases with beam energy, as the core captures more particles (see appendix \ref{sec:fluidization_parameters}).

The proton results are affected by the light nuclei production, which we do not perform, so the comparison to experimental data is not straightforward. In $\sqrtsnn=3\ \GeV$ collisions at STAR, this would account for $\approx21$ protons at midrapidity \cite{STAR:2023uxk}. An excess of protons is expected towards lower energies, which is seen for the cascade and DynFlu results but not for iso-$\tau$. We include the lowest beam energy $\sqrtsnn=2.69\ \GeV$ to highlight this discrepancy. This is a consequence of a significant part of the fluid falling into the so-called fluid corona, i.e.\ regions of the fluid which are outside of particlization surface at the start of the medium evolution. We do not have specific treatment of the corona, so energy and baryon charge of the fluid corona are essentially lost\footnote{In traditional hybrid approaches, the fluid corona is particlized instantaneously at the starting time, which is possible because the corona typically does not enter into the core. Here we use a source-based evolution (for both iso-$\tau$ and DynFlu), and the fluid corona introduced at a given time can become part of the core in the next time step, because of incoming sources. In this case, particlizing the fluid corona would result in double counting its contribution.}. This loss is not significant at higher beam energies and in the dynamic fluidization scenario, where the hydrodynamic evolution receives more energetic and earlier sources. Since the iso-$\tau$ initial condition is anyhow unphysical at these energies, we leave this improvement for the future. 

Midrapidity yields in the DynFlu approach are similar to the cascade evolution, but the forward rapidities have fewer protons until the ``spectator region'', which show higher peaks. This happens because of how core-corona interactions are implemented, i.e. only elastic processes with a reduced total cross section are allowed, causing fewer of the initial nucleons to scatter away from the beam rapidity. 

\begin{figure}[ht]
    \centering
    \includegraphics[width=0.8\linewidth]{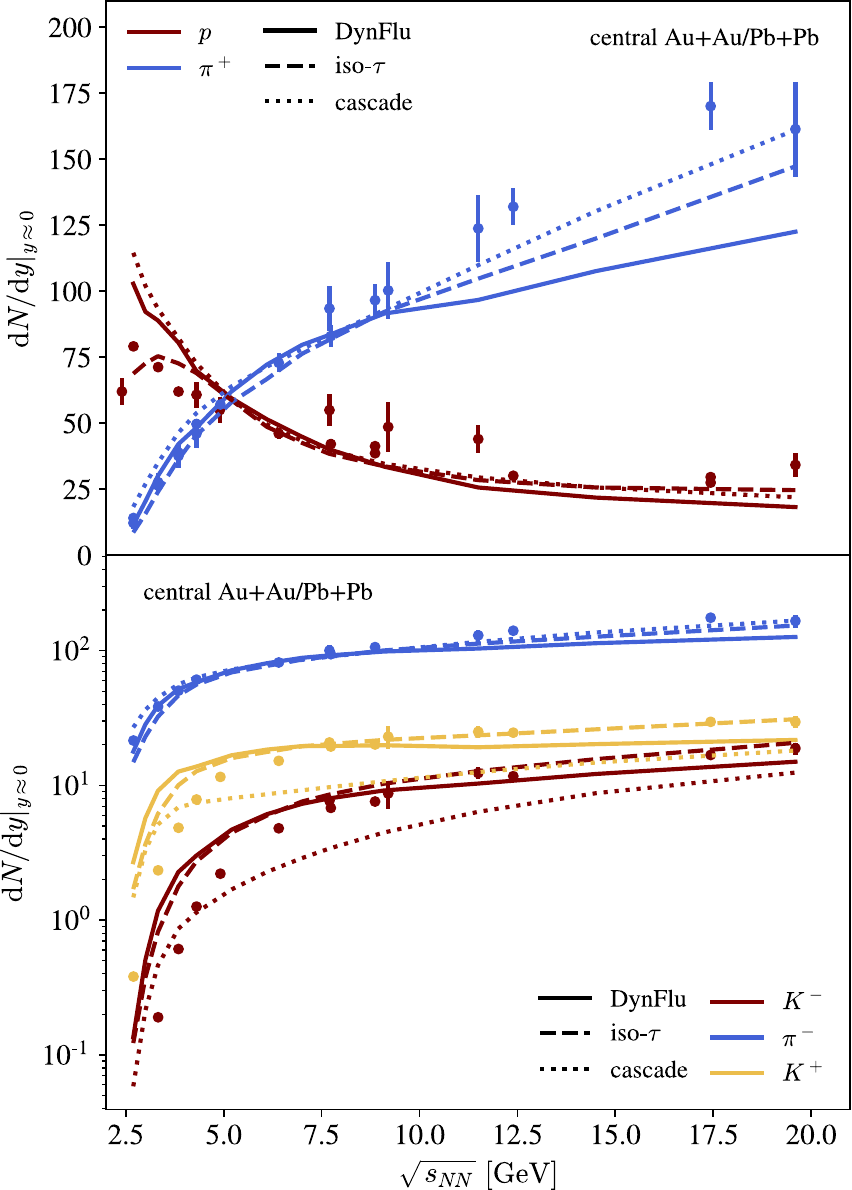}
    \caption{Excitation function of the midrapidity yield of identified particle species for 0-5\% Au+Au collisions, comparing results from the different evolution models. Data points are collected from several experiments \cite{E-0895:2003oas,E895:2001zms,E866:1997cqs,STAR:2021yiu,STAR:2022gki,FOPI:2004bfz}, for both Au+Au and Pb+Pb with varying centrality classes and rapidity cuts.
    }\label{fig:y_excitation}
\end{figure}

Moving towards higher energies, we plot the excitation function of the midrapidity yield for identified particles in fig. \ref{fig:y_excitation}. Between $\sqrtsnn=4.3$ and $9.1\ \GeV$, the dynamic fluidization scenario produces slightly more pions than the iso-$\tau$ initial condition. Above this energy, as already observed in the tuning plots in fig. \ref{fig:tuning_smearing}, the pion yield is severely underestimated. Protons are relatively well described, with a small underproduction at higher energies. The kaons always increase when hydrodynamics is turned on, as strange hadrons are more easily sampled from a thermal distribution than produced from pure hadronic interactions between nucleons. Because of this, the kaon yields are overestimated by the hybrid framework below $7.5\ \GeV$, in particular below $4.3\ \GeV$, where the kaon production in the cascade is already too high.

\subsubsection*{Transverse and elliptic flow}

\begin{figure*}[th]
    \centering    
    \includegraphics[width=0.32\linewidth]{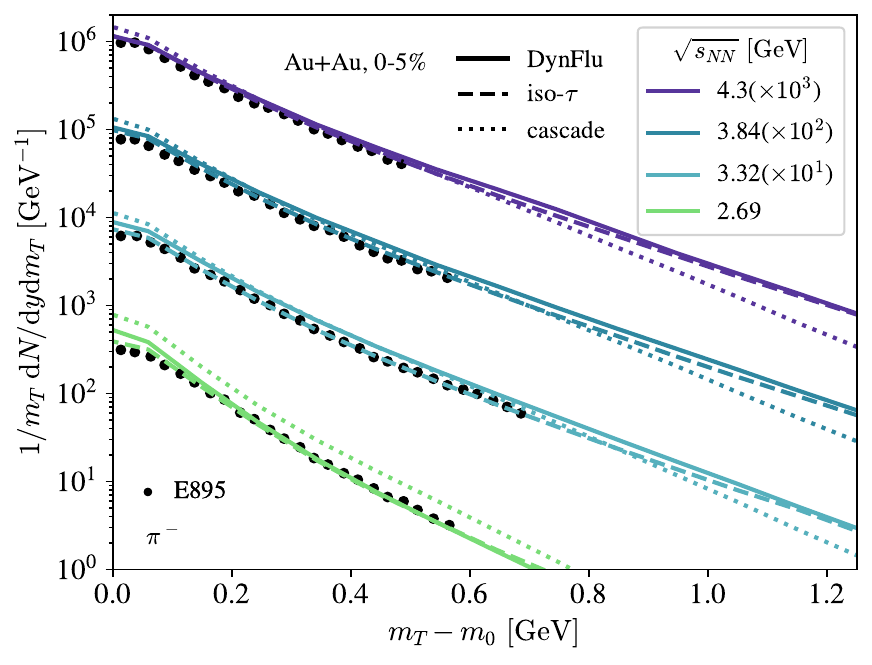}%
    \includegraphics[width=0.32\linewidth]{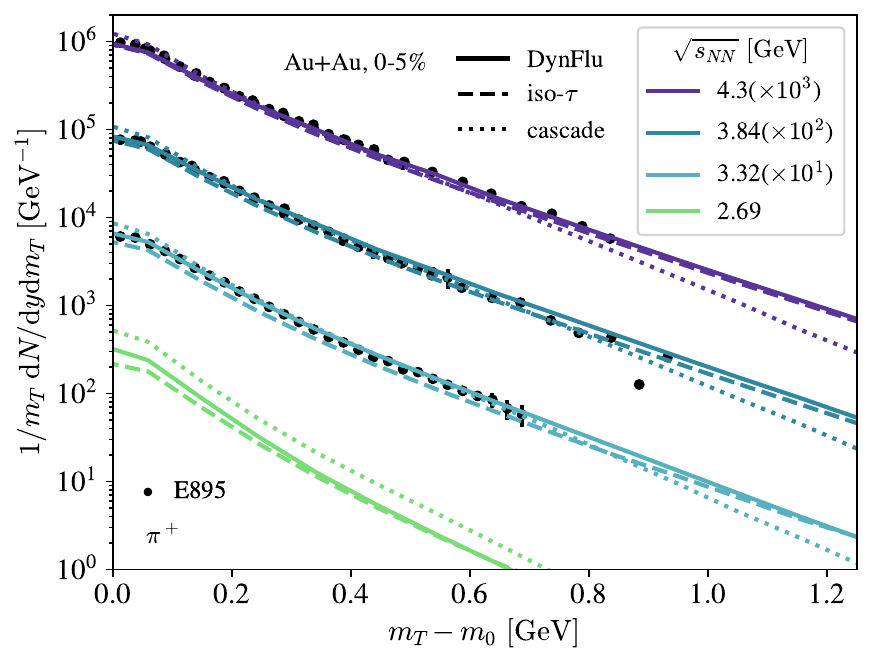}%
    \includegraphics[width=0.33\linewidth]{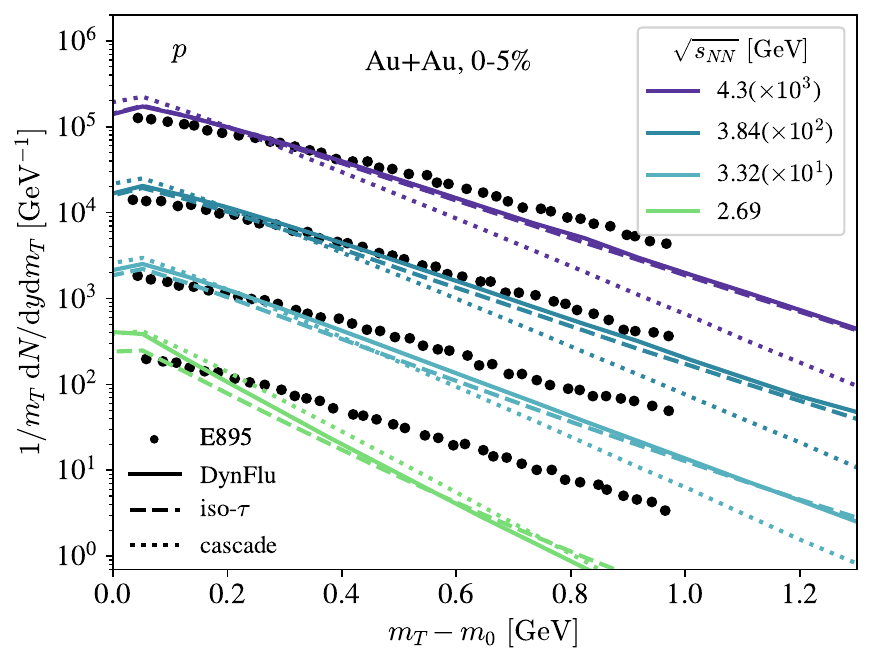}
    \caption{Transverse mass spectra for the same particles and setups as fig. \ref{fig:bulk_y_low_energy}. Data taken from the E895 experiment \cite{E-0895:2003oas}.
    }\label{fig:bulk_mT_low_energy}
\end{figure*}

Figure \ref{fig:bulk_mT_low_energy} shows the respective transverse mass spectra at midrapidity ($|y|<0.1$). The radial flow exhibits no significant differences between DynFlu and iso-$\tau$ initial conditions besides the lowest beam energy, and compared to the cascade evolution, the presence of hydrodynamics induces harder spectra. This improves the match to data, in particular at low $m_T$. The light nuclei again hinder the comparison of proton results, as they are much heavier and thus distort the mass spectra, but this improves with beam energy in the hybrid framework, while spectra remains too soft in a cascade-only scenario.

\begin{figure}[ht]
    \centering
    \includegraphics[width=0.8\linewidth]{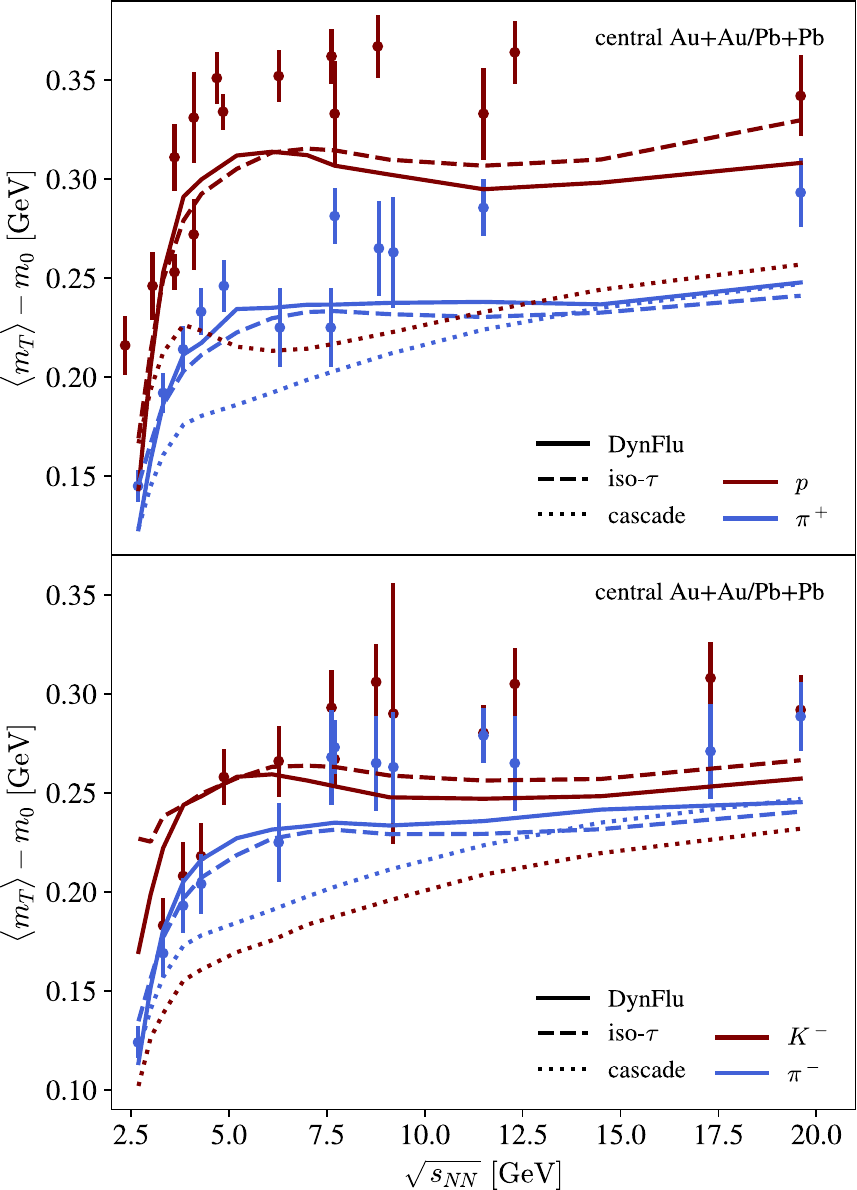}
    \caption{Same as fig. \ref{fig:y_excitation} for the mean transverse mass. The $K^+$ curve is mostly equal to the $K^-$, so we omit it for clarity. Data points are collected from several experiments \cite{E866:1999ktz,NA49:2002pzu,NA49:2004mrq,NA49:2007stj,STAR:2009sxc,STAR:2017sal}, for both Au+Au and Pb+Pb with varying centrality classes and rapidity cuts.}\label{fig:mT_excitation}
\end{figure}

In fig. \ref{fig:mT_excitation}, we show the excitation function of the mean transverse mass for identified particles. Overall, the two hybrid approaches develop similar radial flow, considerably harder than the pure cascade. Although the experimental data is disperse and with large error bars, it is systematically above the dynamic fluidization results by about $50\ \mathrm{MeV}$ at higher energies. The iso-$\tau$ prescription matches the data a bit better, but is still below.

\begin{figure}[ht]
    \centering
    \includegraphics[width=0.9\linewidth]{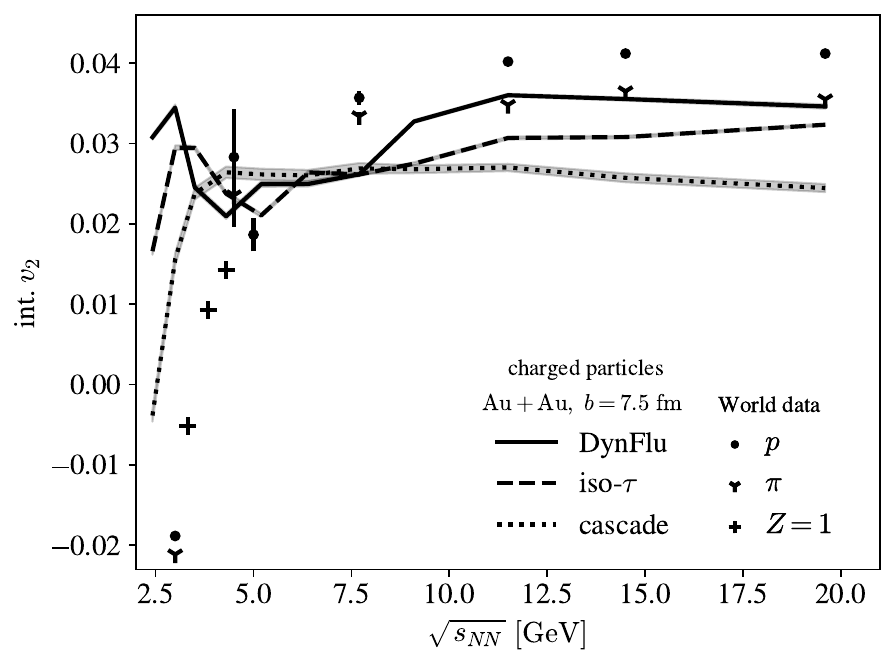}
    \caption{Excitation function of the elliptic flow of charged particles for Au+Au collisions at impact parameter $b=7.5\ \fm$ comparing results from the different evolution models. Experimental data taken from \cite{STAR:2021yiu} and references therein, for identified species in large centrality bins.
    }\label{fig:v2_excitation}
\end{figure}

We show the excitation function of the integrated elliptic flow for charged hadrons in fig. \ref{fig:v2_excitation}, calculated with the reaction plane for particles at midrapidity and $0.2\leq p_T\leq2\ \GeV$. For this, we use a fixed impact parameter $b=7.5\ \fm$, which falls within the centrality class 20-30\% according to Glauber model calculations \cite{nuclearoverlap}. The experimental data comes from a collection of experiments with different ions and definitions of centrality/midrapidity, as such serve only as a reference and we do not aim at a quantitative agreement.

At low energies, the measured $v_2$ is negative due to the \emph{squeeze-out} effect. In the pure hadronic evolution, even without mean-field potentials, the in-plane expansion is blocked by the spectators, so it decreases towards lower collision energies. In contrast, because the hydrodynamic evolution in our framework does not account for spectators, the flow is always positive. At higher beam energies, the pressure gradients in the fluid cause a flow larger than the hadronic cascade, as expected. Compared to the iso-$\tau$ initial condition, in DynFlu the hydrodynamic evolution starts earlier and with a higher energy density, so the fluid has a longer lifetime and generates a larger $v_2$.

The comparisons above showcase that dynamic fluidization successfully extends the SMASH-vHLLE hybrid, providing a proper description of heavy-ion collisions at low beam energies with a hydrodynamic evolution. It is consistent with previous results \cite{Schafer:2021csj}, being equivalent to the original initialization approach based on a constant $\tau$ in the $\sqrtsnn=4.3-9.1\ \GeV$ range.

\subsubsection*{Viscous effects}

The transport coefficients are a major ingredient for the hydrodynamic evolution, therefore, we explore different settings in this section. The ideal version can be compared to the JAM dynamic fluidization approach \cite{Akamatsu:2018olk} and the UrQMD-Shasta hybrid \cite{Steinheimer:2011mp}, while the constant viscosities are the same values used in \cite{Karpenko:2015xea,Schafer:2021csj}. The full parametrization is taken from the Bayesian analysis \cite{Gotz:2025wnv} and represents the setting that has been chosen above.  

\begin{figure}[ht]
    \includegraphics[width=0.9\linewidth]{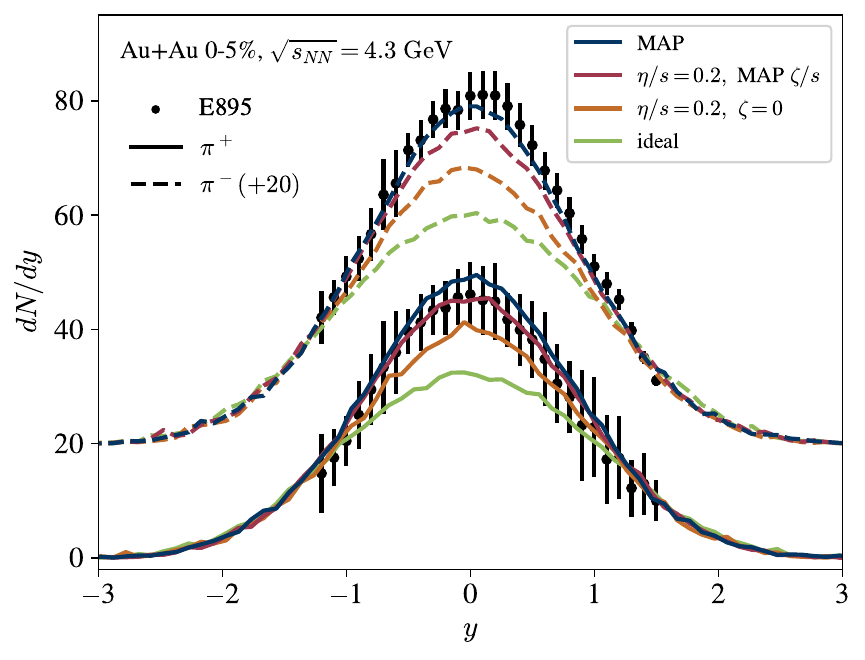}
    
    \includegraphics[width=0.9\linewidth]{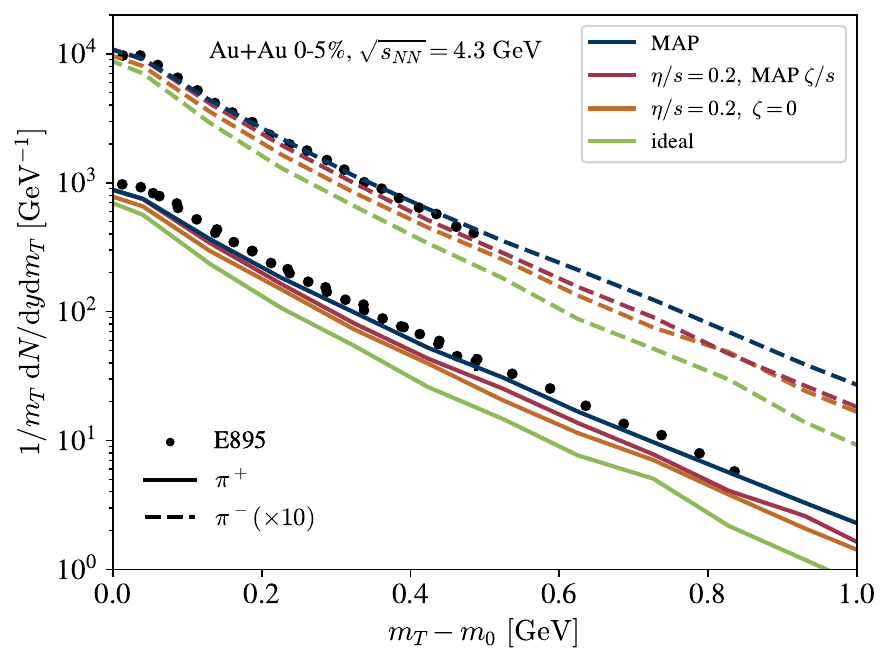}
    \caption{Effects of the viscosity parametrization in (upper) rapidity distributions and (lower) transverse mass spectra of pions in central Au+Au collisions at $\sqrtsnn=4.3\ \mathrm{GeV}$, calculated with dynamic fluidization. Data taken from the E895 experiment \cite{E-0895:2003oas}.}\label{fig:viscous}
\end{figure}

The Bayesian analysis was performed for a different setup of the hybrid framework, so a natural question is whether the \emph{maximum a posteriori} (MAP) set of parameters found is valid for the current dynamic fluidization setup. We check this in fig. \ref{fig:viscous} by changing the viscosity parametrization between the MAP (same curve as DynFlu in figs. \ref{fig:bulk_y_low_energy} and \ref{fig:bulk_mT_low_energy}), a constant value for shear viscosity $\eta/s=0.2$ with or without bulk viscosity $\zeta/s$, and ideal hydrodynamics as in \cite{Akamatsu:2018olk}. The remaining parameters are fixed as discussed in section \ref{sec:tuning}.

Since a viscous system produces entropy, a larger $\eta/s$ leads to a larger pion yield. The MAP parametrization shear viscosity decreases as a function of temperature until some critical temperature $T_c$. At $\sqrtsnn=4.3\ \mathrm{GeV}$, most of the system evolution is below $T_c$, showing an effective viscosity larger than $0.2$. In all curves, the data favors the MAP set of parameters, and the ideal hydrodynamics is the furthest from experiment. 

We take this as indication that the MAP parameters is an adequate base on which we can study and tune the dynamic fluidization model. 

\section{Conclusions and Outlook}\label{sec:discussion}

We have introduced dynamic initial conditions for the hydrodynamic evolution within the SMASH-vHLLE hybrid approach, extending its application regime to lower beam energies. This dynamic fluidization (DynFlu) consists of a core-corona separation implemented in the hadronic transport approach SMASH with two parameters, namely a minimum energy density $\epsth$ to convert hadrons into fluid, and the string fluidization time $t_f$, that determines the quasi-free streaming of string fragmentation products before they become eligible for fluidization. The hadrons are then smeared covariantly with a Gaussian kernel of width $\sigma$ and used as sources for the viscous hydrodynamic evolution in vHLLE, until another energy density limit $\epssw$ is crossed, activating the particlization and afterburner.

This approach is justified by the monotonic decrease of mean free paths with energy density, which was evaluated for a thermalized hadron gas. The characteristic length of the volume defined by $\varepsilon>\epsth$ in the relevant heavy-ion collisions is considerably larger than the mean free paths at $\epsth$, so the interaction rate in this volume is high, hinting at a behavior consistent with hydrodynamics. The \emph{dynamic} character of the initial conditions is needed at lower beam energies due to both spatial and event-wise fluctuations, which are of larger importance regarding the usage of a hydrodynamic description, in particular at the edge of the medium.

The original hybrid framework was developed in \cite{Schafer:2021csj} with initial conditions based on a fixed hyperbolic time $\tau_0$ (iso-$\tau$), and was used down to $\sqrtsnn=4.3\ \GeV$. A set of physical parameters ($\eta$, $\zeta$, and $\epssw$, among others) was determined by a recent Bayesian analysis in \cite{Gotz:2025wnv}. We use these parameters and only tune the newly introduced fluidization time $t_f$ and smearing width $\sigma$. The bulk observables showed little sensitivity to the threshold $\epsth$ within the investigated ranges, which speaks for the equivalence between Boltzmann transport and Israel-Stewart hydrodynamics around $\epsth$.

The DynFlu picture performs well in the bulk observables between $\sqrtsnn=3-9.1\ \GeV$, being in agreement with the iso-$\tau$ initialization above $4.3\ \GeV$. However, it fails for flow observables since the spectators and their squeezing-out are invisible to the hydrodynamic evolution. Above this range, the core-corona separation with the parameters we used becomes too transparent, so the yields at midrapidity are too small compared to experimental data. 

Ideally, the two initialization schemes should converge at higher energies, which can be achieved by requiring that the set of hadrons selected in DynFlu matches the $\tau_0$ hypersurface. Since this highly depends not only on $t_f$, but also on details of string processes and cross sections, such effort is left for the future.

Another improvement would be to run the hadronic cascade and the fluid evolution concurrently, with two-way interaction between both codes. This is quite a challenging procedure, so for the purpose of this first study we mimic it with a technically easier setup, where the fluid evolution does not have a back reaction on the hadronic cascade. The only exception are the hadrons sampled from the fluid at particlization and added to the final-state hadronic cascade, together with the corona hadrons that never fluidized. Such parallel evolution will also allow for appropriate calculations of anisotropic flow, by taking the spectator nucleons into account.

In this work we have neglected nuclear potentials, which play an important role for pure hadronic transport at low beam energy collisions. We assume their effect decreases once a hydrodynamic evolution is included, but this will be checked in a future study.

Since the recent predictions for the QCD critical point fall within the range where dynamic fluidization describes the evolution well, it will be particularly interesting to explore the consequences of different equations of state that include first-order phase transitions.

Furthermore, electromagnetic probes are very sensitive to the initial dynamics of the expansion, as they are emitted early on. While the yields of final state hadrons at low energies are similar between the dynamic fluidization and iso-$\tau$ initializations, the yields of dileptons will be rather different, in particular at the intermediate mass range. As such, coupling the hydrodynamic evolution with thermal emission rates is a promising next step.

\section*{Acknowledgments}

This work was supported by the Helmholtz Forschungsakademie Hessen für FAIR (HFHF). The authors also acknowledge the support by the State of Hesse within the Research Cluster ELEMENTS (Project ID 500/10.006), and by the Deutsche Forschungsgemeinschaft (DFG, German Research Foundation) – Project number 315477589 – TRR 211. IK acknowledges support by the Czech Science Foundation under project No. 25-16877S. Computational resources have been provided by the GreenCube at GSI. 

\appendix

\section{Mean free paths in a hadron gas}\label{sec:mfp}

A hydrodynamic description can be applied instead of kinetic theory for the evolution of a system if the interaction rate is large enough \cite{landau1987fluid}. With particle degrees of freedom, a natural probe for this is the average distance particles travel between interactions, the mean free path $\lambda$. If it is considerably smaller than the typical medium scale, we expect that the Boltzmann equation is equivalent to hydrodynamics \cite{Niemi:2014wta}, justifying the use of a hybrid framework. In a heavy-ion collision, defining this average on an event-by-event basis is ambiguous, so instead we use SMASH to extract the mean free path of nucleons, pions, and kaons in a thermalized hadron resonance gas with $\mu_B=0$. By doing so, we can map the microscopic reaction dynamics to a mean energy density that allows for the assumption of local equilibration to be reasonable and therefore serve as a fluidization criterion. 

\begin{figure}[t]
\centering
\includegraphics[width=.85\linewidth]{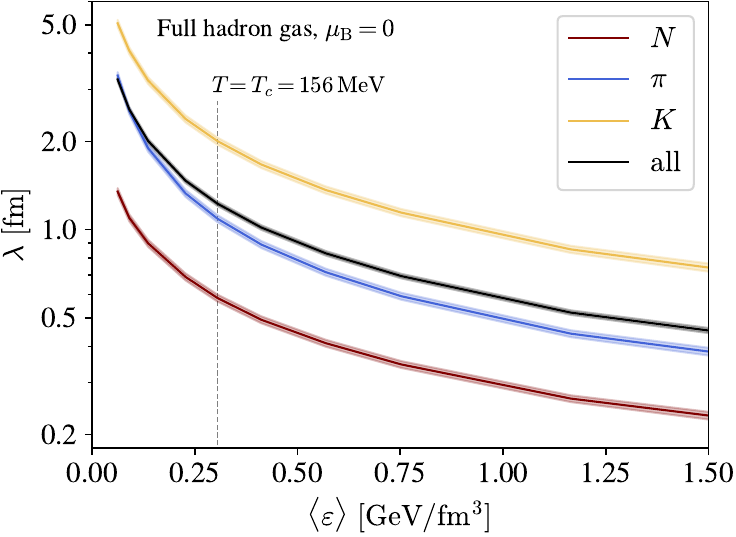}\caption{Mean free path of nucleons, pions, kaons, and all hadrons as a function of energy density in an equilibrated hadron gas. The energy density corresponding to the crossover temperature $T_c$, determined by lattice QCD \cite{HotQCD:2018pds}, is highlighted.} \label{fig:mean_free_path}
\end{figure}

In fig. \ref{fig:mean_free_path}, we show that $\lambda$ decreases with energy density, as expected from the classic kinetic theory relation $\lambda\sim(n\sigma)^{-1}$, where $n$ is the particle density and $\sigma$ is the interaction cross section. For the latter, SMASH uses the Additive Quark Model \cite{Goulianos:1982vk,Bass:1998ca}, where mesons and strange hadrons have reduced cross sections compared to nucleons, leading to the hierarchy $\lambda_N<\lambda_\pi<\lambda_K$. Because of the large number of other resonances which have a small cross section, the multiplicity averaged $\lambda_\mathrm{all}$ ends up about the same as $\lambda_\pi$. In a low energy heavy-ion collision, however, the medium is much more nucleon abundant, which would bring this curve down. The monotonic dependence of these curves advocates for the local energy density as a probe of mean free paths, in particular in off-equilibrium scenarios where $\lambda$ cannot be directly estimated, such as heavy-ion collisions.

\section{Bounds for the fluidization parameters}\label{sec:fluidization_parameters}

\begin{figure*}[ht]
\centering
\includegraphics[height=0.34\linewidth]{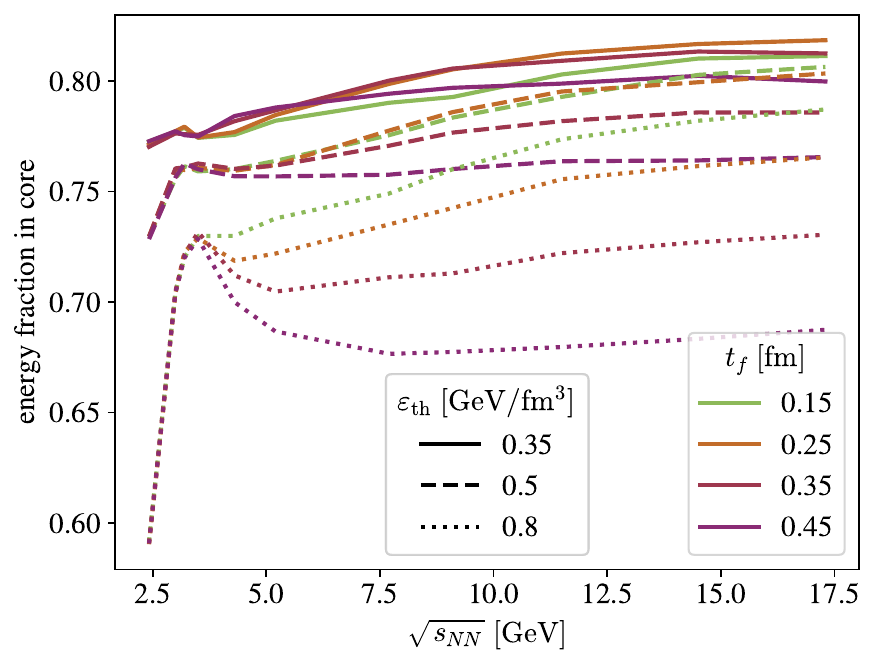}%
\includegraphics[height=0.34\linewidth]{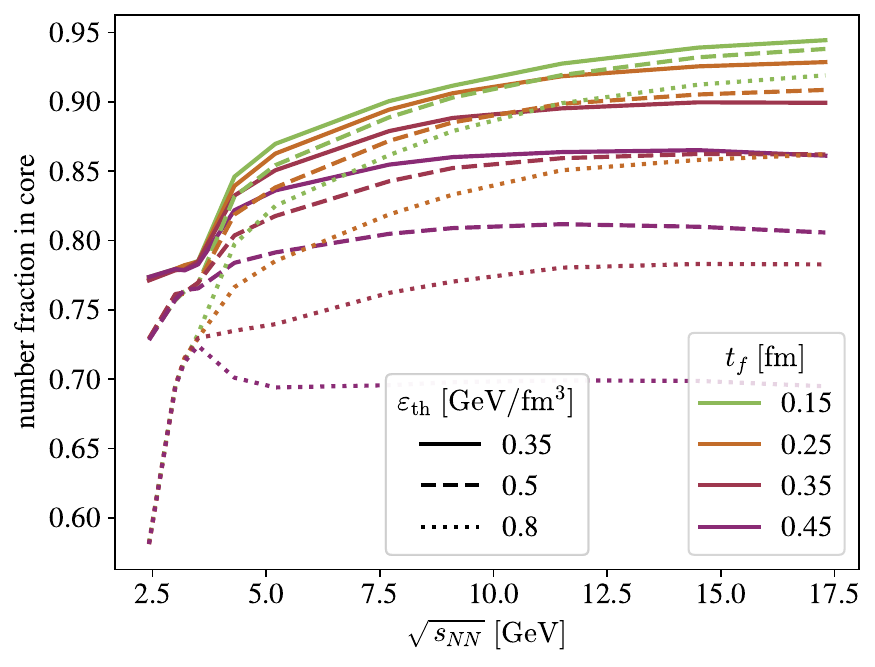}
\caption{Fraction of the total energy (left) and number of particles (right) given to the core as a function of the beam energy in central Pb+Pb collisions for different fluidization parameters.}\label{fig:ratio_total_core}
\end{figure*}

The core-corona procedure described in the main text introduces two parameters, the threshold energy density and the string fluidization time. Their effect on the core-corona separation is summarized in fig. \ref{fig:ratio_total_core}: $\epsth$ parametrizes how \emph{difficult} it is to locally change from a hadronic description into a hydrodynamical one, as such the deposition into the core is higher for lower, more relaxed values. $t_f$ determines the delay between the creation of a string and the deposition of its content into the core, and hence it only affects collisions with $\sqrtsnn>3.5\ \GeV$, when the string cross section is nonzero. Although physically motivated, they represent an effective description of hydrodynamization, and there is no first-principles calculation wherefrom to obtain them. Still, we can impose physical bounds that prevent undesirable behavior in the initial state.

Since $\epsth$ controls the change in the physical description from hadrons to fluid, the naïve assumption would be to fix it to the switching energy density $\epssw$, which determines the reverse change at the end of the hydrodynamic evolution. However, for particlization the system starts in (or close to) equilibrium, unlike in the fluidization, where we neglect that particles may be far off-equilibrium. A hadron resonance gas approaches thermalization faster when the energy density is higher \cite{Rose:2020lfc}, so we require $\epsth\geq\epssw$. This means that the translation particles $\to$ fluid should be more difficult than the reverse. Moreover, the partonic evolution post-initial scatterings happens entirely in the hydrodynamic stage, so the fluidization must happen below the energy densities that characterize Quark-Gluon Plasma formation. Therefore, we restrict the threshold to the bounds
\begin{equation}
    \epssw\leq\epsth < \varepsilon_\mathrm{QGP},
\end{equation}
with $\epssw=0.334\ \GeVfm^3$ and $\varepsilon_\mathrm{QGP}\approx1.0\ \GeVfm^3$. In this range, we found little to no sensitivity on bulk observables for which there is available data.

Moreover, $t_f$ must be non-zero, otherwise fluidization becomes instantaneous for the string products, as the energy needed to create a string guarantees that the local energy density will exceed $\epsth$. On the other hand, the fraction of the medium that gets deposited into the core should  be larger at higher beam energies. This imposes a simultaneous constraint in $\epsth$ and $t_f$. For example, as we show in Fig. \ref{fig:ratio_total_core}, for $\epsth=0.5\ \GeVfm^3$ we require the limits
\begin{equation}
    0<t_f \leq 0.45\ \fm,
\end{equation}
with the caveat that this upper bound depends on details of the cross sections for string processes.

\section{Viscosity parameters}\label{sec:visc_parameters}

Here we summarize the viscosity parametrizations used in the Bayesian analysis performed in \cite{Gotz:2025wnv}, along with the maximum a posteriori estimates for the parameters.

The shear viscosity is given by
\begin{equation}
    \frac{\eta T}{\varepsilon + P} = \left(1 + a_{\mu_B}\frac{\mu_B}{\mu_{B,0}}\right) f(T-T_c),\quad 
\end{equation}
where 
\begin{equation}
f(x)= \max \left( 0,\left(\frac{\eta}{s}\right)_{\text{min}}  +\label{eq:shear} \begin{cases}
a_{l,\eta}x,\quad x < 0\\
a_{h,\eta}x,\quad x \geq 0\\
\end{cases}\right),
\end{equation}
and
\begin{equation}
T_c = T_{\eta,0} + b_{\mu_B} \frac{\mu_B}{\mu_{B,0}}.
\end{equation}

The bulk viscosity is parametrized as
\begin{equation}
    \frac{\zeta T}{\varepsilon + P} = \zeta_0 \exp\left(-\beta\frac{(\varepsilon^{1/4}-\varepsilon_\zeta^{1/4})^2}{2\sigma_{\zeta}^2}\right),
\end{equation}
incorporating the $\mu_B$ dependence implicitly on the energy density. The Gaussian width $\sigma_\zeta$ is asymmetric: 
\begin{equation}
\sigma_\zeta=\begin{cases}
\sigma_{\zeta,-} & \varepsilon < \varepsilon_\zeta, \\
\sigma_{\zeta,+} & \varepsilon > \varepsilon_\zeta,
\end{cases} 
\end{equation}

The corresponding parameters are listed in table \ref{tab:map_parameter}:
\begin{table}[h!]
    \centering
    \renewcommand{\arraystretch}{1.5}
    \begin{tabular}{c|c c}
        \hline \hline
        \textbf{Parameter} & \textbf{MAP} &\\ \hline
        $a_{l,\eta}$           & $-10.9149^{+6.5179}_{-3.2712}$ & $\GeV^{-1}$ \\ 
        $a_{h,\eta}$           & $-10.5116^{+3.3021}_{-3.3842} $& $\GeV^{-1}$\\ 
        $T_{\eta,0}$           & $0.1480^{+0.0223}_{-0.0435}$ & GeV  \\ 
        $(\eta/s)_{\text{min}}$ & $0.2214^{+0.0806}_{-0.0931}$ & \\ 
        $a_{\mu_B}$     & $2.8399^{+3.8013}_{-2.4225}$ & \\ 
        $b_{\mu_B}$     & $0.1379^{+0.7536}_{-0.3114} $& \\ 
         $\zeta_0$           & $0.0816^{+0.0785}_{-0.0483} $& \\ 
        $\varepsilon_\zeta$    & $24.0359^{+11.7138}_{-13.3645} $& $\GeVfm^3$ \\ 
        $\sigma_{\zeta,-}$  & $0.0725^{+0.0154}_{-0.0474}$ & \\ 
        $\sigma_{\zeta,+}$  &$ 0.0378^{+0.0913}_{-0.0025} $& \\ \hline \hline
    \end{tabular}
    \caption{MAP of all parameters of the model.}
    \label{tab:map_parameter}
\end{table}

\bibliography{references}

\end{document}